\documentclass[5p]{elsarticle}
\pdfoutput=1
\usepackage{relsize}
\usepackage{etoolbox}
\newtoggle{submission}
\toggletrue{submission}
\newtoggle{TwoColumnFormulas}
\toggletrue{TwoColumnFormulas}
\biboptions{longnamesfirst,square,sort,compress}

\usepackage[english]{babel}
\usepackage[utf8]{inputenc}
\usepackage[T1]{fontenc}   
\usepackage{graphicx}
\usepackage{enumerate}

\usepackage{color,amssymb,amsmath}
\usepackage{txfonts}
\usepackage{empheq}
\usepackage{mathtools}
\usepackage{mathrsfs} 
\DeclareMathAlphabet{\mathscrbf}{OMS}{mdugm}{b}{n} 
\usepackage{frcursive}
\usepackage{comment}
\usepackage{booktabs}
\usepackage[font=footnotesize,labelfont=bf,tableposition=top]{caption}

\usepackage{placeins}

\usepackage{tikz}

\usepackage{pgfplots}
\iftoggle{submission}{}{
  \usetikzlibrary{pgfplots.groupplots}
  \usetikzlibrary{calc}
  \pgfplotsset{compat=1.12}
  \usepgfplotslibrary{external}
  \tikzexternalize
}

\newlength{\GraphWidth}
\newlength{\GraphHeight}
\newlength{\GraphHorizSep}
\newlength{\GraphVerticSep}
\newlength{\GraphVerticLegendSep}
\newlength{\HorizVerticLegendSep}
\makeatletter
\if@twocolumn
  \tikzset{every mark/.append style={scale=1.2,line width=0.3pt}}
  \pgfplotsset{legend image code/.code={\draw[mark repeat=3, mark phase=2,#1] plot coordinates {(0cm,0cm)(0.3cm,0cm)(0.6cm,0cm)};\draw[color=white] (0cm,-0.07cm) rectangle (0.6cm,0.07cm);}}
  \newenvironment{SmartFigure}{\begin{figure}}{\end{figure}}
  \newcommand{\MyLWone}{0.75pt}  
  \newcommand{\MyLWtwo}{1pt} 
\else
  \tikzset{every mark/.append style={line width=0.6pt}}
  \pgfplotsset{legend image code/.code={\draw[mark repeat=3, mark phase=2,#1] plot coordinates {(0cm,0cm)(0.5cm,0cm)(1.0cm,0cm)};\draw[color=white] (0cm,-0.1cm) rectangle (1.0cm,0.1cm);}}
  \newenvironment{SmartFigure}{\begin{figure*}}{\end{figure*}}
  \newcommand{\MyLWone}{1pt}  
  \newcommand{\MyLWtwo}{2pt} 
\fi
\pgfplotscreateplotcyclelist{MyCycleListOne}{{black,only marks,mark=star,mark options={scale=1.3},mark repeat={10}},{blue,densely dotted,line width = \MyLWone},{green,only marks,mark=o,mark repeat={10}},{red,loosely dotted,line width = \MyLWtwo},{black}}
\pgfplotscreateplotcyclelist{MyCycleListTwo}{{green,line width = \MyLWone},{black,loosely dotted,line width = \MyLWone},{blue,densely dotted,line width = \MyLWone},{red,loosely dotted,line width = \MyLWtwo},{black}}
\makeatother

\usepackage{float}
\restylefloat{table}

\usepackage{cancel}

\newcommand{\mymat}[1]{
[\mathbf{#1}]}
\newcommand{\myvect}[1]{
\{\mathbf{#1}\}}

\allowdisplaybreaks[4]

\specialcomment{myproofs}
{\begingroup\color{blue}}{\endgroup}
\specialcomment{Tomodify}
{\begingroup\color{red}}{\endgroup}
\excludecomment{myproofs}

\usepackage{hyperref}
\hypersetup{
    unicode=true,          
    pdftoolbar=true,        
    pdfmenubar=true,        
    pdffitwindow=true,     
    pdfstartview={FitH},    
    pdftitle={A multilayered plate theory with transverse shear and normal warping functions},    
    pdfauthor={Alexandre Loredo},     
    pdfcreator={Alexandre Loredo},   
    pdfproducer={Alexandre Loredo}, 
    pdfnewwindow=true,      
    colorlinks=true,       
    linkcolor=red,          
    citecolor=green,        
    filecolor=magenta,      
    urlcolor=cyan           
}

\newcommand{\WFfirst}{(WFs)}
\newcommand{\WF}{WF}
\newcommand{\WFs}{WFs}
\newcommand{\tozzfour}{ToZZ\nobreakdash-4}
\newcommand{\sizzfour}{SiZZ\nobreakdash-4}
\newcommand{\tdfourwf}{3D\nobreakdash-4WF}
\newcommand{\tozzfive}{ToZZ\nobreakdash-5}
\newcommand{\sizzfive}{SiZZ\nobreakdash-5}
\newcommand{\tdfivewf}{3D\nobreakdash-5WF}

\input cyracc.def

\journal{Composite Structures}

\begin{document}

\relscale{0.9}

\begin{frontmatter}
  \title{A multilayered plate theory with transverse shear and normal warping functions}
	\author[aff]{A.~Loredo\corref{cor1}}
	\ead{alexandre.loredo@u-bourgogne.fr}
	\cortext[cor1]{Corresponding author}
	\address[aff]{DRIVE -- Universit\'e de Bourgogne, 49 rue Mlle Bourgeois, 58027 Nevers, France}
	\begin{abstract}
A multilayered plate theory taking into account transverse shear and normal stretching is presented. The theory is based on a seven-unknowns kinematic field with five warping functions. Four warping functions are related to the transverse shear, the fifth to the normal stretching. The warping functions are issued from exact three-dimensional solutions. They are related to the variations of transverse shear and normal stresses computed at specific points for a simply supported bending problem. Reddy, Cho--Parmerter and (a modified version of) Beakou--Touratier theories have been retained for comparisons. Extended versions of these theories, able to manage the normal stretching, are also considered. These theories, which use the same kinematic field with different warping functions, are compared to analytical solutions for the bending of simply supported plates. Various plates are considered, with special focus on low length-to-thickness ratios: an isotropic plate, two homogeneous orthotropic plates with ply orientation of $0$ and $5$ degrees, a $[0/c/0]$ sandwich panel and a $[-45/0/45/90]_s$ composite plate. Results show that models are more accurate if their kinematic fields (i) depend on all material properties (not only the transverse shear stiffnesses) (ii) depend on the length-to-thickness ratios (iii) present a coupling between the $x$ and $y$ directions.
	\end{abstract}
	\begin{keyword}
	  Plate theory \sep warping function \sep normal stretching \sep laminate \sep sandwich \sep vibration
	\end{keyword}
\end{frontmatter}
	\newpage
\section{Introduction}
Plate theories have been enhanced in order to model structures becoming more and more complex over the years. For thin homogeneous plates, works of Cauchy~\cite{Cauchy1828}, Kirchhoff~\cite{Kirchhoff1850} and Love~\cite{Love1888} have lead to the so-called Love--Kirchhoff theory, which does not take into account the transverse shear. For moderately thick homogeneous plates, transverse shear must appear in the formulation. Authors like Reissner, Hencky, Bolle, Uflyand, Hildebrand and Mindlin~\cite{Reissner1945, Hencky1947, Bolle1947, Uflyand1948, Hildebrand1949, Mindlin1951} have proposed to integrate the shear phenomenon into their formulation. In particular, Reissner made assumptions on stresses, hence shear stresses have a parabolic distribution and the normal stress is considered in his model which is derived from a complementary energy. The other authors made assumptions on displacements. For example, Hencky and Bolle considered a linear variation of the displacements $u$ and $v$ with constant transverse shear strains, but no normal strain. Mindlin developed the dynamic version of Hencky's theory. Mindlin's and Reissner's theories are often associated but it is incorrect as demonstrated in reference~\cite{Wang2001}. The Hencky--Mindlin theory tends to overestimate the transverse shear stiffnesses. Hence, a shear correction factor has been proposed, generally fixed to the value of $5/6$ for static studies of homogeneous plates. Except works such as L\'evy's memoir~\cite{Levy1877}, Hildebrand's second-order theory~\cite{Hildebrand1949} and Vlasov's and Murthy 's third-order theories~\cite{Vlasov1957,Murthy1981}, higher-order theories have mainly been proposed for inhomogeneous structures, and hence are presented below. The models presented in the 1877 Lévy’s memoir were ahead of their time, with a displacement field which depends on $z$ through the sum of a polynomial of degree $3$ and a sine function.
\par
Laminated composite plates, including sandwich panels, are an important class of structures, widely used in the industrial field. The mechanical behaviour of such structures is difficult to model in the general case, because of their heterogeneous nature. Although previous works on heterogeneous plates and/or sandwiches have been done by authors like Lekhnitskii~\cite{Lekhnitskii1941}, Reissner~\cite{Reissner1947, Reissner1950} among others, the Classical Lamination Theory, which is the multilayer extension of the Love--Kirchhoff theory is generally atributed to Stravsky~\cite{Stavsky1961}, Reissner and Stavsky~\cite{Reissner_1961}, and Dong \& al.~\cite{Dong1962}. The First order Shear Deformation laminated plate Theory (FoSDT) based on the Hencky--Mindlin theory is attributed to Yang \& al.~\cite{Yang1966} and Whitney~\cite{Whitney1969, Whitney1970}. The use of shear correction factors is here mandatory because the overestimation of the FoSDT for shear stiffnesses is even worse for laminated and sandwich structures than for homogeneous ones. Although everybody agree for the value of $5/6$ for the shear correction factor for homogeneous plates in static studies, values for the 3 needed correction factors of general laminates may take different values depending on the method used to calculate them~\cite{Whitney1973,Noor1989,Pai1995} and it has been proved that they depend on the wavelength~\cite{Chatterjee1979}.
\par
It is possible to model laminated plates with a layerwise approach, but this leads to a number of unknowns which depends on the number of layers, which may be large. This class of layerwise (LW) theories is opposed to the class of equivalent single layer (ESL) theories which contains all the previously cited theories. In the ESL class, theories have a more or less great number of unknowns, but which does not depend on the number of layers.
\par
In this ESL framework, the complex behaviour of laminated structures has pushed researchers into proposing higher order plate theories. These plate theories are characterized by the use a displacement field with a higher order (higher than one) dependence on the normal coordinate $z$. Except the work of Whitney~\cite{Whitney1973b} which presents a second order theory, the well known and commonly used higher order theories are of order $3$ like the Levinson's~\cite{Levinson1980} and Reddy's~\cite{Reddy1984} ones. Other higher order theories have been proposed in following works, which differ on the unknowns which are considered, the order of the developments, etc.~\cite{Carrera2003}. Non polynomial theories have also been proposed, characterized by the use of trigonometric, hyperbolic or such similar functions of $z$ to model the displacement field~\cite{Touratier1991,Soldatos1992,Thai2014}. They are often considered and classified as higher order theories.
\par
Some \emph{a priori} LW models can reduce to ESL models with the help of assumptions between the fields in each layer. Zig-Zag (ZZ) models enter in this category. Early works of Lekhnitskii~\cite{Lekhnitskii1935} and Ambartsumyan~\cite{Ambartsumyan1958} have been classified as such by Carrera~\cite{Carrera2002} who shows also that other authors have integrated the multilayer structure in their model~\cite{Whitney1969, Sun1973, Cho1993} in a very similar manner. The main idea of ZZ models is to let in-plane displacements vary with $z$ according to the superposition of a zig-zag law to a global law -- cubic for example. With these models, shear stresses can satisfy both continuity at interfaces and null (or prescribed) values at the top and bottom faces of the plate.
\par
In reference~\cite{Loredo2014b}, authors present a plate model with four \emph{warping functions} \WFfirst{} $\varphi_{\alpha\beta}(z)$ which embed the transverse shear behaviour into the displacement field. These functions are issued, for each lamination sequence, from 3D elasticity solutions. An important conclusion of this paper is that, as all ``without-normal-stretching'' models use reduced stiffnesses according to the generalized plane stress assumption, the comparison of these models with an exact 3D solution has no sense for very low length-to thickness ratios like $2$ or $4$. The results provided by these models can however be compared to an exact solution for virtual laminates which have been highly stiffened in the $z$ direction, see the paper~\cite{Loredo2014b} for more details. The corollary is that, to compare models with 3D solutions at very low length-to-thickness ratios, models must integrate the normal stretching behaviour.
\par
Plate theories taking into account the normal stretching behaviour have been proposed since a long time~\cite{Levy1877,Reissner1950} and in many other works as it can be seen in recent reviews on the subject~\cite{Wanji2008,Kreja2011,Khandan2012}. However, there is a recent interest for them, motivated by the need to accurate simulations of sandwich panels~\cite{Pai2001} and functionally graded materials. In reference~\cite{Mantari2014b} a trigonometric plate theory with normal stretching is developed. It presents an interesting mechanism to adapt the slopes of the shear and normal \WFs{}, with a choice depending on the length-to-thickness ratio, which illustrate the need to such adaptation. In the present paper, as in the corresponding previous work~\cite{Loredo2014b}, we use \WFs{} that depends on length-to-thickness ratios because they are issued from 3D elasticity solutions. In the same reference~\cite{Mantari2014b}, there is no coupling between the $x$ and $y$ directions in the kinematic. Further, this theory belongs in a particular class of theories which consider a splitting of the transverse displacement into two parts $w=w_b+w_s$, where $w_b$ and $w_s$ are respectively the bending and shear contributions to deflection. This splitting seems to have been first introduced to avoid problems with the clamped boundary condition for Reddy's like third-order models, as it is explained in reference~\cite{Murty1987}. In this reference, in addition to the $w_s$ unknown, independent shear strain unknowns are considered. Although this splitting remains interesting, using it without independent shear strain unknowns, as it is done in reference~\cite{Mantari2014b} and other similar works, leads (according to formulas 2d--e and 3h--m of reference~\cite{Mantari2014b}) to $\gamma_{xz,y}=\gamma_{yz,x}$ which may be restrictive. In reference\cite{Barut2013} a refined zig-zag model called RZT \{2,2\} is used to study cross-ply laminates with a high number of layers including adhesive layers and resin-rich layers. The model has $11$ unknown functions that do not correspond to the $7$ of the present paper, and therefore cannot easily be implemented for comparison. In addition, no angle-ply, nor general lamination scheme are studied and the model has no coupling between the $x$ and $y$ direction in its kinematic.
\par
The present theory is an extension of one of the two models presented in reference~\cite{Loredo2014b}, aiming to take into account the normal deformation. Two supplementary unknowns and a fifth \WF{} $\varphi_{33}(z)$ are used to model the stretching phenomenon. The theory is compared to other theories and to exact solutions.
\section{Considered plate theory}\label{sec:Formulation}
\subsection{Laminate definition and index convention}
The laminate, of height $h$, is composed of $N$ layers. All the quantities are related to unknown functions defined at the $z=0$ middle plane, and which are marked with the superscript $0$.
%
In the following, Greek subscripts take values $1$ or $2$ and Latin subscripts take values $1$, $2$ or $3$. Einstein's summation convention is used for subscripts only. The comma used as a subscript index means the partial derivative with respect to the following indices.
\subsection{Displacement, strain and stress fields}\label{sec:DisplacementField}
The kinematic assumptions of the theory are
\begin{subequations}
\label{eq:depl}
\begin{empheq}[left=\empheqlbrace]{align}
    u_\alpha(x,y,z) & =  u_\alpha^0(x,y) - z w^0_{,\alpha}(x,y) + \varphi_{\alpha\beta}(z)\gamma^0_{\beta3}(x,y) \label{eq:in_plane_depl} \\ 
    u_3(x,y,z) & =  w^0(x,y) + z \varepsilon^0_{33}(x,y) + \varphi_{33}(z)\kappa^0_{33}(x,y) \label{eq:normal_depl}
\end{empheq}
\end{subequations}
where $u_\alpha^0(x,y)$, $w^0(x,y)$ are the in-plane displacements and the deflection evaluated at $z=0$, $\gamma^0_{\alpha3}(x,y)$ are the engineering transverse shear strains evaluated at $z=0^-$, $\varepsilon^0_{33}(x,y)$ and $\kappa^0_{33}(x,y)$ are the values of the first and second derivatives of $w(x,y,z)$ with respect to $z$ at $z=0^-$, and $\varphi_{\alpha\beta}(z)$ and $\varphi_{33}(z)$ are the five \WFs{}, which are continuous. The unknowns which are evaluated at $z=0^-$ correspond to quantities which may be discontinuous if an interface between different materials is located at the middle plane $z=0$. The above definitions imply $\varphi_{\alpha\beta}(0)=0$, $\varphi_{33}(0)=\varphi'_{33}(0^-)=0$ and $\varphi''_{33}(0^-)=1$.
The associated strain field is derived from equation~\eqref{eq:depl}:
\iftoggle{TwoColumnFormulas}{
  \begin{subequations}\label{eq:strains}
  \begin{empheq}[left=\empheqlbrace]{align}
    \varepsilon_{\alpha\beta}(x,y,z) & = \varepsilon_{\alpha\beta}^0(x,y) - z w^0_{,\alpha\beta}(x,y) + \frac{1}{2}\Big(\varphi_{\alpha\gamma}(z)\gamma^0_{\gamma3,\beta}(x,y)
        \iftoggle{TwoColumnFormulas}{\nonumber \\&}{} 
                                       + \varphi_{\beta\gamma}(z)\gamma^0_{\gamma3,\alpha}(x,y)\Big) \label{eq:in_plane_strains}\\ 
    \varepsilon_{\alpha3}(x,y,z) & = \frac{1}{2}\Big(\varphi'_{\alpha\beta}(z)\gamma^0_{\beta3}(x,y) + z\varepsilon^0_{33,\alpha}(x,y)
        \iftoggle{TwoColumnFormulas}{\nonumber \\&}{} 
                                   + \varphi_{33}(z)\kappa^0_{33,\alpha}(x,y)\Big) \label{eq:transverse_strains}\\ 
    \varepsilon_{33}(x,y,z) & = \varepsilon^0_{33}(x,y) + \varphi'_{33}(z)\kappa^0_{33}(x,y)\phantom{\frac{1}{2}}\label{eq:normal_strain}
  \end{empheq}
  \end{subequations}
}{
  \begin{subequations}\label{eq:strains}
  \begin{empheq}[left=\empheqlbrace]{align}
    \varepsilon_{\alpha\beta}(x,y,z) & = \varepsilon_{\alpha\beta}^0(x,y) - z w^0_{,\alpha\beta}(x,y) + \frac{1}{2}\Big(\varphi_{\alpha\gamma}(z)\gamma^0_{\gamma3,\beta}(x,y)
                                       + \varphi_{\beta\gamma}(z)\gamma^0_{\gamma3,\alpha}(x,y)\Big) \label{eq:in_plane_strains}\\ 
    \varepsilon_{\alpha3}(x,y,z) & = \frac{1}{2}\Big(\varphi'_{\alpha\beta}(z)\gamma^0_{\beta3}(x,y) + z\varepsilon^0_{33,\alpha}(x,y)
                                   + \varphi_{33}(z)\kappa^0_{33,\alpha}(x,y)\Big) \label{eq:transverse_strains}\\ 
    \varepsilon_{33}(x,y,z) & = \varepsilon^0_{33}(x,y) + \varphi'_{33}(z)\kappa^0_{33}(x,y)\phantom{\frac{1}{2}}\label{eq:normal_strain}
  \end{empheq}
  \end{subequations}
}
In addition to the aforementioned conditions, equation~\eqref{eq:transverse_strains} shows that the \WFs{} must also verify $\varphi'_{\alpha\beta}(0^-)=\delta_{\alpha\beta}$.

Applying Hooke's law leads to the following stress field ($x$ and $y$ are omitted for clarity):
\begin{subequations}\label{eq:stresses}
\begin{empheq}[left=\empheqlbrace]{align}
  \sigma_{\alpha\beta}(z) & = C_{\alpha\beta\gamma\delta}(z)\left(\varepsilon_{\gamma\delta}^0 - z w^0_{,\gamma\delta} + \varphi_{\gamma\mu}(z)\gamma^0_{\mu3,\delta}\right) 
    \iftoggle{TwoColumnFormulas}{\nonumber \\&}{} 
                            + C_{\alpha\beta33}(z)\left(\varepsilon^0_{33} + \varphi'_{33}(z)\kappa_{33}^0\right)\label{eq:in_plane_stresses} \\ 
  \sigma_{\alpha3}(z) & = C_{\alpha3\beta3}(z)\left(\varphi'_{\beta\mu}(z)\gamma^0_{\mu3} + z\varepsilon^0_{33,\beta} +\varphi_{33}(z)\kappa^0_{33,\beta} \right)\label{eq:transverse_stresses2}\\ 
  \sigma_{33}(z) & = C_{33\alpha\beta}(z)\left(\varepsilon_{\alpha\beta}^0 - z w^0_{,\alpha\beta} + \varphi_{\alpha\mu}(z)\gamma^0_{\mu3,\beta}\right)
      \iftoggle{TwoColumnFormulas}{\nonumber \\&}{} 
                   + C_{3333}(z)\left(\varepsilon^0_{33} + \varphi'_{33}(z)\kappa_{33}^0\right)
\end{empheq}
\end{subequations}
\subsection{Strain energy, generalized forces and strains}\label{sec:StrainEnergy}
Let us now consider the strain energy surface density:
\begin{myproofs}
\begin{align}\label{eq:StrainEnergy1p}
  J&=\frac{1}{2}\int_{-h/2}^{h/2}\varepsilon_{ij}\sigma_{ij}\text{d}z
	  =\frac{1}{2}\int_{-h/2}^{h/2}\left(\varepsilon_{\alpha\beta}\sigma_{\alpha\beta}+2\varepsilon_{\alpha3}\sigma_{\alpha3}+\varepsilon_{33}\sigma_{33}\right)\text{d}z \nonumber \\
   &=\frac{1}{2}\int_{-h/2}^{h/2}
	  \bigg[\Big(\varepsilon_{\alpha\beta}^0 - z w^0_{,\alpha\beta} + \frac{1}{2}\left(\varphi_{\alpha\gamma}(z)\gamma^0_{\gamma3,\beta}
		  +\varphi_{\beta\gamma}(z)\gamma^0_{\gamma3,\alpha}\right)\Big)\sigma_{\alpha\beta} \nonumber \\
		&\phantom{=\frac{1}{2}\int_{-h/2}^{h/2}\bigg[ }+2\times\frac{1}{2}\left(\varphi'_{\alpha\beta}(z)\gamma^0_{\beta3} + z\varepsilon^0_{33,\alpha} +\varphi_{33}(z)\kappa^0_{33,\alpha}\right)\sigma_     {\alpha3}
   \iftoggle{TwoColumnFormulas}{\nonumber \\&\phantom{=\frac{1}{2}\int_{-h/2}^{h/2}\bigg[ }}{} 
     +\left(\varepsilon^0_{33} + \varphi'_{33}(z)\kappa_{33}^0\right)\sigma_{33}\bigg]\text{d}z \nonumber \\
   &=\frac{1}{2}\int_{-h/2}^{h/2}
	  \Big[\left(\varepsilon_{\alpha\beta}^0 - z w^0_{,\alpha\beta} + \varphi_{\alpha\gamma}(z)\gamma^0_{\gamma3,\beta}\right)\sigma_{\alpha\beta}   
   \iftoggle{TwoColumnFormulas}{\nonumber \\&\phantom{=\frac{1}{2}\int_{-h/2}^{h/2}\bigg[ }}{} 
     +\left(\varphi'_{\alpha\beta}(z)\gamma^0_{\beta3} + z\varepsilon^0_{33,\alpha} +\varphi_{33}(z)\kappa^0_{33,\alpha}\right)\sigma_{\alpha3}
   \iftoggle{TwoColumnFormulas}{\nonumber \\&\phantom{=\frac{1}{2}\int_{-h/2}^{h/2}\Big[ }}{} 
     +\left(\varepsilon^0_{33} +\varphi'_{33}(z)\kappa_{33}^0\right)\sigma_{33}\Big]\text{d}z
\end{align}
\end{myproofs}
\begin{align}\label{eq:StrainEnergy1}
  J&=\frac{1}{2}\int_{-h/2}^{h/2}\varepsilon_{ij}\sigma_{ij}\text{d}z
	  =\frac{1}{2}\int_{-h/2}^{h/2}\left(\varepsilon_{\alpha\beta}\sigma_{\alpha\beta}+2\varepsilon_{\alpha3}\sigma_{\alpha3}+\varepsilon_{33}\sigma_{33}\right)\text{d}z \nonumber \\
   &=\frac{1}{2}\int_{-h/2}^{h/2}
	  \Big[\left(\varepsilon_{\alpha\beta}^0 - z w^0_{,\alpha\beta} + \varphi_{\alpha\gamma}(z)\gamma^0_{\gamma3,\beta}\right)\sigma_{\alpha\beta}
    \iftoggle{TwoColumnFormulas}{\nonumber \\&\phantom{=\frac{1}{2}\int_{-h/2}^{h/2}\Big[ }}{} 
    +\left(\varphi'_{\alpha\beta}(z)\gamma^0_{\beta3} + z\varepsilon^0_{33,\alpha} +\varphi_{33}(z)\kappa^0_{33,\alpha}\right)\sigma_{\alpha3}
    \iftoggle{TwoColumnFormulas}{\nonumber \\&\phantom{=\frac{1}{2}\int_{-h/2}^{h/2}\Big[ }}{} 
    +\left(\varepsilon^0_{33} +\varphi'_{33}(z)\kappa_{33}^0\right)\sigma_{33}\Big]\text{d}z
\end{align}
%
%
It can also be written
\begin{align}\label{eq:StrainEnergy21}
    J =\frac{1}{2}&
	  \Big[\varepsilon_{\alpha\beta}^0 N_{\alpha\beta} + \kappa^0_{\alpha\beta} M_{\alpha\beta} + \gamma^0_{\gamma3,\beta} P_{\gamma\beta} + \gamma^0_{\alpha3} N_{\alpha3} 
            \iftoggle{TwoColumnFormulas}{\nonumber \\&}{} 
          + \kappa^0_{\alpha3}M_{\alpha3} + \kappa^0_{33,\alpha}P_{\alpha3} + \varepsilon^0_{33}N_{33} + \kappa^0_{33}M_{33} \Big]
\end{align}
where $\kappa^0_{\alpha\beta}=-w^0_{,\alpha\beta}$ and $\kappa^0_{\alpha3}=\varepsilon^0_{33,\alpha}$, naturally introducing the $18$ following quantities which are the generalized forces,
\begin{subequations}\label{eq:generalized_forces1}
\begin{empheq}[left=\empheqlbrace]{align}
  \{N_{\alpha\beta},M_{\alpha\beta},P_{\gamma\beta}\}&=\int_{-h/2}^{h/2} \{1,z,\varphi_{\alpha\gamma}(z)\} \sigma_{\alpha\beta}(z) \text{d}z \\
  \{N_{\beta3}, M_{\alpha3}, P_{\alpha3}\}&=\int_{-h/2}^{h/2} \{\varphi'_{\alpha\beta}(z), z,\varphi_{33}(z) \} \sigma_{\alpha3}(z) \text{d}z \\
  \{N_{33},M_{33}\} &=\int_{-h/2}^{h/2} \{1, \varphi'_{33}(z) \} \sigma_{33}(z) \text{d}z 
\end{empheq}
\end{subequations}
each associated with a corresponding generalized displacement in the strain energy formula~\eqref{eq:StrainEnergy21}.
They are then set, by type, into vectors
\begin{align}\label{eq:generalized_forces_v3}
  &\mathbf{N}=
  \begin{Bmatrix}
    N_{11} \\
		N_{22} \\
		N_{12} \\
		N_{33}
  \end{Bmatrix}
  \quad
  \mathbf{M}=
  \begin{Bmatrix}
    M_{11} \\
		M_{22} \\
	  M_{12} \\
    M_{33}
  \end{Bmatrix}
  \quad
  \mathbf{P}=
  \begin{Bmatrix}
    P_{11} \\
		P_{22} \\
		P_{12} \\
	  P_{21} \\
  \end{Bmatrix}
  \iftoggle{TwoColumnFormulas}{\nonumber \\&}{\quad} 
  \mathscrbf{N}=
  \begin{Bmatrix}
    N_{13} \\
		N_{23}
  \end{Bmatrix}
	\quad
  \mathscrbf{M}=
  \begin{Bmatrix}
    M_{13} \\
		M_{23}
  \end{Bmatrix}
	\quad  
  \mathscrbf{P} =
  \begin{Bmatrix}
    P_{13} \\
		P_{23}
  \end{Bmatrix}
\end{align} 
and the same is done for the corresponding generalized strains:
\begin{align}\label{eq:generalized_strains_v3}
  &\boldsymbol{\varepsilon}=
  \begin{Bmatrix}
    \varepsilon^0_{11} \\
		\varepsilon^0_{22} \\
		2 \varepsilon^0_{12} \\
		\varepsilon^0_{33}
  \end{Bmatrix}
  \quad
  \boldsymbol{\kappa}=
  \begin{Bmatrix}
    \kappa^0_{11} \\
		\kappa^0_{22} \\
	  2\kappa^0_{12} \\
 	  \kappa^0_{33}
  \end{Bmatrix}
  \quad
  \mathbf{\Gamma}=
  \begin{Bmatrix}
    \gamma^0_{13,1} \\
		\gamma^0_{23,2} \\
		\gamma^0_{13,2} \\
	  \gamma^0_{23,1} 
  \end{Bmatrix}
  \iftoggle{TwoColumnFormulas}{\nonumber \\&}{\quad} 
  \boldsymbol{\gamma}=
  \begin{Bmatrix}
    \gamma^0_{13} \\
		\gamma^0_{23}
  \end{Bmatrix}
  \quad
  \boldsymbol{\lambda}=
  \begin{Bmatrix}
    \kappa^0_{13} \\
		\kappa^0_{23}
  \end{Bmatrix}
  \quad
  \boldsymbol{\mu}=
  \begin{Bmatrix}
    \kappa^0_{33,1} \\
		\kappa^0_{33,2}
  \end{Bmatrix}
\end{align}
\subsection{Laminate behaviour}\label{sec:LaminateBehaviour}
Generalized forces are linked with the generalized strains by the $12\times12$ and $6\times6$ following stiffness matrices
\begin{equation}\label{eq:behavior_v2}
\begin{Bmatrix}
  \mathbf{N} \\
	\mathbf{M} \\
	\mathbf{P}
\end{Bmatrix}
=
\begin{bmatrix}
  \mathbf{A}   & \mathbf{B}   & \mathbf{E} \\
	\mathbf{B}   & \mathbf{D}   & \mathbf{F} \\
	\mathbf{E^T} & \mathbf{F^T} & \mathbf{G} 
\end{bmatrix}
\begin{Bmatrix}
  \boldsymbol{\varepsilon} \\
	\boldsymbol{\kappa} \\
	\boldsymbol{\Gamma} \\
\end{Bmatrix}
\quad\text{and}\quad
\begin{Bmatrix}
  \mathscrbf{N} \\
  \mathscrbf{M} \\
  \mathscrbf{P}
\end{Bmatrix}
=
\begin{bmatrix}
  \mathbf{H} & \mathbf{I} & \mathbf{J} \\
  \mathbf{I}^\mathbf{T} & \mathbf{K} & \mathbf{L} \\
  \mathbf{J}^\mathbf{T} & \mathbf{L}^\mathbf{T} & \mathbf{O}
\end{bmatrix}
\begin{Bmatrix}
	\boldsymbol{\gamma} \\
	\boldsymbol{\lambda} \\
	\boldsymbol{\mu}
\end{Bmatrix}
\end{equation}
with the following definitions
\begin{align}\label{eq:generalized stiffnesses1_v3}
  \{A_{\alpha\beta\gamma\delta},B_{\alpha\beta\gamma\delta},D_{\alpha\beta\gamma\delta},E_{\alpha\beta\mu\delta},F_{\alpha\beta\mu\delta},G_{\nu\beta\mu\delta}\}
  \iftoggle{TwoColumnFormulas}{}{&}
  =\int_{-h/2}^{h/2} C_{\alpha\beta\gamma\delta}(z)
    \iftoggle{TwoColumnFormulas}{\ast\nonumber \\}{}
    \{1,z,z^2,\varphi_{\gamma\mu}(z),z\varphi_{\gamma\mu}(z),\varphi_{\alpha\nu}(z)\varphi_{\gamma\mu}(z)\}\text{d}z \nonumber\\
  \{A_{\alpha\beta33},B_{\alpha\beta33},B_{33\alpha\beta},D_{\alpha\beta33},E_{33\gamma\beta},F_{33\gamma\beta}\}
  \iftoggle{TwoColumnFormulas}{}{&}
  =\int_{-h/2}^{h/2} C_{33\alpha\beta}(z)
    \iftoggle{TwoColumnFormulas}{\ast\nonumber \\}{}
    \{1,\varphi'_{33}(z),z,z\varphi'_{33}(z),\varphi_{\alpha\gamma}(z),z\varphi_{\alpha\gamma}(z)\}\text{d}z \nonumber\\
  \{A_{3333},B_{3333},D_{3333}\}
  \iftoggle{TwoColumnFormulas}{}{&}
  =\int_{-h/2}^{h/2} C_{3333}(z)\{1,\varphi'_{33}(z),\varphi'^2_{33}(z)\}\text{d}z \nonumber\\
  \{H_{\alpha3\beta3},I_{\alpha3\delta3},J_{\gamma3\delta3},K_{\alpha3\beta3},L_{\alpha3\delta3},O_{\gamma3\delta3} \}
  \iftoggle{TwoColumnFormulas}{}{&}
  =\int_{-h/2}^{h/2} C_{\gamma3\delta3}(z)
    \iftoggle{TwoColumnFormulas}{\ast\nonumber \\}{}
  \{\varphi'_{\gamma\alpha}(z)\varphi'_{\delta\beta}(z),z\varphi'_{\gamma\alpha}(z),\varphi'_{\gamma\alpha}(z)\varphi_{33}(z),z^2,z\varphi_{33}(z), \varphi^2_{33}(z)\} \text{d}z
\end{align}
and:
\begin{align}\label{eq:generalized stiffnesses2}
  A_{33\alpha\beta}=A_{\alpha\beta33} \quad ; \quad D_{33\alpha\beta}=D_{\alpha\beta33}
\end{align}
Taking into account symmetries, this leads to (in the order which their appear in formulas~\eqref{eq:generalized stiffnesses1_v3}) $6+6+6+12+12+10+3+3+3+3+4+4+1+1+1+3+4+4+3+3+3=95$ independent stiffnesses in the more general case.
\subsection{Kinetic energy}
With the help of the displacement field expressions~\eqref{eq:depl}, the kinetic energy surface density $E_c(x,y)$ of the structure can be written
\begin{myproofs}
\begin{align}\label{eq:KineticEnergy1_proof}
  E_c(x,y)&=\frac{1}{2}\int_{-h/2}^{h/2} \rho(x,y,z)\dot{u}_{i}(x,y,z)\dot{u}_{i}(x,y,z)\text{d}z \nonumber \\
	        &=\frac{1}{2}\int_{-h/2}^{h/2} \rho(z)\bigg[\left(\dot{u}_\alpha^0 - z \dot{w}^0_{,\alpha} + \varphi_{\alpha\beta}(z)\dot{\gamma}^0_{\beta3}\right) \left(\dot{u}_\alpha^0 
          - z \dot{w}^0_{,\alpha} + \varphi_{\alpha\beta}(z)\dot{\gamma}^0_{\beta3}\right) 
          \iftoggle{TwoColumnFormulas}{\nonumber \\&\phantom{=\frac{1}{2}\int_{-h/2}^{h/2}\rho(z)\bigg[}}{}
          + \left( \dot{w}^0 + z \dot{\varepsilon}^0_{33} + \varphi_{33}(z)\dot{\kappa}^0_{33}\right)^2\bigg]\text{d}z \nonumber \\
	        &=\frac{1}{2}\int_{-h/2}^{h/2} \rho(z)\bigg[\dot{u}^0_\alpha\dot{u}^0_\alpha - 2 z \dot{u}^0_\alpha \dot{w}^0_{,\alpha} + 2 \varphi_{\alpha\beta}(z) \dot{u}^0_\alpha \dot{\gamma}^0_{\beta3}
          + z^2 \dot{w}^0_{,\alpha}\dot{w}^0_{,\alpha} 
          \iftoggle{TwoColumnFormulas}{\nonumber \\&\phantom{=\frac{1}{2}\int_{-h/2}^{h/2}\rho(z)\bigg[}}{}
          - 2 z \varphi_{\alpha\beta}(z) \dot{w}^0_{,\alpha} \dot{\gamma}^0_{\beta3} + \varphi_{\alpha\beta}(z)\varphi_{\alpha\mu}(z)\dot{\gamma}^0_{\beta3} \dot{\gamma}^0_{\mu3} \nonumber \\
		      &\phantom{&=\frac{1}{2}\int_{-h/2}^{h/2}\rho(z)\bigg[} + (\dot{w}^0)^2 + z^2(\dot{\varepsilon}^0_{33})^2 + \varphi^2_{33}(z)(\dot{\kappa}^0_{33})^2 + 2 z \dot{w}^0\dot{\varepsilon}^0_{33} 
          \iftoggle{TwoColumnFormulas}{\nonumber \\&\phantom{=\frac{1}{2}\int_{-h/2}^{h/2}\rho(z)\bigg[}}{}
          + 2\varphi_{33}(z)\dot{w}^0\dot{\kappa}^0_{33} + 2z\varphi_{33}(z)\dot{\varepsilon}^0_{33}\dot{\kappa}^0_{33} \bigg]\text{d}z
\end{align}
\end{myproofs}
\begin{align}\label{eq:KineticEnergy1}
  E_c(x,y)&=\frac{1}{2}\int_{-h/2}^{h/2} \rho(x,y,z)\dot{u}_{i}(x,y,z)\dot{u}_{i}(x,y,z)\text{d}z \nonumber \\
          &=\frac{1}{2}\Big(R \dot{u}^0_\alpha \dot{u}^0_\alpha - 2 S \dot{u}^0_\alpha \dot{w}^0_{,\alpha} + 2 U_{\alpha\beta} \dot{u}^0_\alpha \dot{\gamma}^0_{\beta3} 
          + T \dot{w}^0_{,\alpha}\dot{w}^0_{,\alpha} 
          \iftoggle{TwoColumnFormulas}{\nonumber \\&\phantom{\frac{1}{2}\Big(R}}{}
          - 2 V_{\alpha\beta} \dot{w}^0_{,\alpha} \dot{\gamma}^0_{\beta3} + W_{\alpha\beta} \dot{\gamma}^0_{\alpha3} \dot{\gamma}^0_{\beta3} 
          \iftoggle{TwoColumnFormulas}{}{\nonumber \\&\phantom{\frac{1}{2}\Big(R}}
          + R(\dot{w}^0)^2 + T(\dot{\varepsilon}^0_{33})^2 
          \iftoggle{TwoColumnFormulas}{\nonumber \\&\phantom{\frac{1}{2}\Big(R}}{}
          + \mathscr{W}(\dot{\kappa}^0_{33})^2 + 2 S \dot{w}^0\dot{\varepsilon}^0_{33} + 2 \mathscr{U} \dot{w}^0\dot{\kappa}^0_{33} + 2 \mathscr{V}\dot{\varepsilon}^0_{33}\dot{\kappa}^0_{33}\Big)
\end{align}
where the following generalized mass have been considered:
\begin{align}\label{eq:generalized_mass}
  \{R,S,T,U_{\alpha\beta},V_{\alpha\beta},W_{\alpha\beta},\mathscr{U},\mathscr{V},\mathscr{W}\} = \int_{-h/2}^{h/2} \rho(z)
  \iftoggle{TwoColumnFormulas}{\ast\nonumber \\}{}
  \{1,z,z^2,\varphi_{\alpha\beta}(z),\varphi_{\alpha\beta}(z)z,\varphi_{\mu\alpha}(z)\varphi_{\mu\beta}(z),\varphi_{33}(z),z\varphi_{33}(z),\varphi^2_{33}(z)\} \text{d}z
\end{align} 
Note that the $U_{\alpha\beta}$ and $V_{\alpha\beta}$ are antisymmetric tensors but $W_{\alpha\beta}$ is symmetric. Then, there are $17$ independent mass coefficients to consider.
\subsection{Laminate equations of motion}\label{sec:LaminateEquationsOfMotion}
Let us recall the equilibrium conditions within a solid. Without loss of generality, body forces are neglected here, and the previous convention on indices is kept:
\begin{subequations}\label{eq:equilibrium0}
\begin{empheq}[left=\empheqlbrace]{align}
  \sigma_{\alpha\beta,\beta}+\sigma_{\alpha3,3}=\rho \ddot{u}_{\alpha} \label{eq:equilibrium0_membrane} \\
  \sigma_{\alpha3,\alpha}+\sigma_{33,3}=\rho \ddot{u}_{3} \label{eq:equilibrium0_normal}
\end{empheq}
\end{subequations} 
\par
Integrating the equations of equilibrium~\eqref{eq:equilibrium0} over the thickness with the help of formulas~\eqref{eq:depl}, \eqref{eq:generalized_forces1} and~\eqref{eq:generalized_mass} leads to
\begin{myproofs}
	{\begin{subequations}\label{eq:equilibrium1demo}
	\begin{empheq}[left=\empheqlbrace]{align}
		\int_{-h/2}^{h/2} \sigma_{\alpha\beta,\beta}\text{d}z+\int_{-h/2}^{h/2} \sigma_{\alpha3,3}\text{d}z
		&=\int_{-h/2}^{h/2} \rho \left[\ddot{u}^0_\alpha - z \ddot{u}^0_{3,\alpha} + \varphi_{\alpha\beta}\ddot{\gamma}^0_{\beta3}\right]\text{d}z \\
		\int_{-h/2}^{h/2} \sigma_{\alpha3,\alpha}\text{d}z + \int_{-h/2}^{h/2} \sigma_{33,3}\text{d}z
		&=\int_{-h/2}^{h/2} \rho \left[\ddot{w}^0 + z\ddot{\varepsilon}^0_{33}(x,y) + \varphi_{33}(z)\ddot{\kappa}^0_{33}(x,y)\right]\text{d}z
	\end{empheq}
	\end{subequations}} 
\end{myproofs}
\begin{subequations}\label{eq:equilibrium1}
\begin{empheq}[left=\empheqlbrace]{align}
  N_{\alpha\beta,\beta}+[\sigma_{\alpha3}(z)]_{-h/2}^{h/2}&=R \ddot{u}^0_{\alpha} - S \ddot{w}^0_{,\alpha}+U_{\alpha\beta}\ddot{\gamma}^0_{\beta3} \\
  Q^c_{\alpha,\alpha}+[\sigma_{33}(z)]_{-h/2}^{h/2}&=R \ddot{w}^0+S\ddot{\varepsilon}^0_{33}+\mathscr{U}\ddot{\kappa}^0_{33} \label{eq:equilibrium1b}
\end{empheq}
\end{subequations} 
where the $Q^c_{\alpha}$ are the classical shear forces. In order to get more equations, weighted integrals over the thickness of equation~\eqref{eq:equilibrium0_membrane} with weight functions $z$ and $\varphi_{\alpha\gamma}(z)$, and of equation~\eqref{eq:equilibrium0_normal} with weight functions $z$ and $\varphi_{33}(z)$ are computed. It gives six more equations:%
\begin{myproofs}
	{\begin{subequations}\label{eq:equilibrium2demo}
	\begin{empheq}[left=\empheqlbrace]{align}
		\int_{-h/2}^{h/2} \sigma_{\alpha\beta,\beta}z\text{d}z+\int_{-h/2}^{h/2} \sigma_{\alpha3,3}z\text{d}z
		&=\int_{-h/2}^{h/2} \rho \left[\ddot{u}^0_\alpha - z \ddot{u}^0_{3,\alpha} + \varphi_{\alpha\beta}\ddot{\gamma}^0_{\beta3}\right]z\text{d}z \\
		\int_{-h/2}^{h/2} \varphi_{\alpha\gamma}\sigma_{\alpha\beta,\beta}\text{d}z+\int_{-h/2}^{h/2} \varphi_{\alpha\gamma}\sigma_{\alpha3,3}\text{d}z
		&=\int_{-h/2}^{h/2} \rho \varphi_{\alpha\gamma}\left[\ddot{u}^0_\alpha - z \ddot{u}^0_{3,\alpha} + \varphi_{\alpha\beta}\ddot{\gamma}^0_{\beta3}\right]\text{d}z\\
		\int_{-h/2}^{h/2} \sigma_{\alpha3,\alpha}z\text{d}z + \int_{-h/2}^{h/2} \sigma_{33,3}z\text{d}z
		&=\int_{-h/2}^{h/2} \rho \left[\ddot{w}^0 + z\ddot{\varepsilon}^0_{33}(x,y) + \varphi_{33}(z)\ddot{\kappa}^0_{33}(x,y)\right]z\text{d}z\\
		\int_{-h/2}^{h/2} \varphi_{33}\sigma_{\alpha3,\alpha}\text{d}z + \int_{-h/2}^{h/2} \varphi_{33}\sigma_{33,3}\text{d}z
		&=\int_{-h/2}^{h/2} \rho \varphi_{33}\left[\ddot{w}^0 + z\ddot{\varepsilon}^0_{33}(x,y) + \varphi_{33}(z)\ddot{\kappa}^0_{33}(x,y)\right]\text{d}z
	\end{empheq}
	\end{subequations}} 
\end{myproofs}
\begin{subequations}\label{eq:equilibrium2}
\begin{empheq}[left=\empheqlbrace]{align}
 & M_{\alpha\beta,\beta}+[\sigma_{\alpha3}(z)z]_{-h/2}^{h/2}-Q^c_{\alpha} = S \ddot{u}^0_{\alpha} - T \ddot{w}^0_{,\alpha}+V_{\alpha\beta}\ddot{\gamma}^0_{\beta3} \label{eq:equilibrium2a}\\
 & P_{\gamma\beta,\beta}+[\varphi_{\alpha\gamma}(z)\sigma_{\alpha3}(z)]_{-h/2}^{h/2}-N_{\gamma3} = U_{\alpha\gamma} \ddot{u}^0_{\alpha} - V_{\alpha\gamma} \ddot{w}^0_{,\alpha}+W_{\gamma\beta}\ddot{\gamma}^0_{\beta3} \\
 & M_{\alpha3,\alpha}+[z\sigma_{33}(z)]_{-h/2}^{h/2}-N_{33} = S \ddot{w}^0 + T\ddot{\varepsilon}^0_{33} + \mathscr{V}\ddot{\kappa}^0_{33}\\
 & P_{\alpha3,\alpha}+[\varphi_{33}(z)\sigma_{33}(z)]_{-h/2}^{h/2}-P_{33} = \mathscr{U} \ddot{w}^0 + \mathscr{V}\ddot{\varepsilon}^0_{33}+ \mathscr{W}\ddot{\kappa}^0_{33} 
\end{empheq}
\end{subequations} 
Let $\{q,r,s\}=[\sigma_{33}(z)\{1,z,\varphi_{33}(z)\}]_{-h/2}^{h/2}$ denote the values of the transverse loading and associated moments, and suppose there is no tangential forces on the top and bottom of the plate, so $\sigma_{\alpha3}(-h/2)=\sigma_{\alpha3}(h/2)=0$. We shall note that there is no generalized strains corresponding to the classical shear forces $Q^c_{\alpha}$. They must be eliminated. It is done by setting values of $Q^c_{\alpha}$ obtained from formula~\eqref{eq:equilibrium2a} into equation~\eqref{eq:equilibrium1b}. This leads to the plate equilibrium system of equations (7 equations with 7 unknown functions):
\begin{subequations}\label{eq:equilibrium3}
\begin{empheq}[left=\empheqlbrace]{align}
  &N_{\alpha\beta,\beta}=R \ddot{u}^0_{\alpha} - S \ddot{w}^0_{,\alpha}+U_{\alpha\beta}\ddot{\gamma}^0_{\beta3} \\
  &M_{\alpha\beta,\beta\alpha} + q =R \ddot{w}^0 + S \ddot{u}^0_{\alpha,\alpha} + S \ddot{\varepsilon}^0_{33} - T \ddot{w}^0_{,\alpha\alpha} + \mathscr{U} \ddot{\kappa}^0_{33} + V_{\alpha\beta}\ddot{\gamma}^0_{\beta3,\alpha} \\
  &P_{\alpha\beta,\beta}-N_{\alpha3}=U_{\beta\alpha} \ddot{u}^0_{\beta} - V_{\beta\alpha} \ddot{w}^0_{,\beta}+W_{\alpha\beta}\ddot{\gamma}^0_{\beta3} \\
  &M_{\alpha3,\alpha} + r - N_{33} = S \ddot{w}^0 + T\ddot{\varepsilon}^0_{33} + \mathscr{V}\ddot{\kappa}^0_{33}\\
  &P_{\alpha3,\alpha} + s - P_{33} = \mathscr{U} \ddot{w}^0 + \mathscr{V}\ddot{\varepsilon}^0_{33}+ \mathscr{W}\ddot{\kappa}^0_{33} 
\end{empheq}
\end{subequations}
\section{Warping functions issued from an exact 3D solution}\label{sec:WF}
The set of \WFs{} is issued from the 3D analytical solution of the bending of a simply supported rectangular plate submitted to a bi-sine load (see section~\ref{sec:Exa} for details). For dynamic studies, the static solution is replaced by the response of the plate to a bi-sine load at a given frequency. Let us define three points of the middle plane A $(\tfrac{a}{2},0)$, B $(0,\tfrac{b}{2})$ and C $(\tfrac{a}{2},\tfrac{b}{2})$, where $a$ and $b$ are the side lengths of the plate. The two-step procedure is described in the following.
\subsection{Computation of the $\varphi_{33}(z)$ warping function}
\label{sec:NormalWF}
The through-the-thickness variations of the normal strain $\varepsilon_{33}$ and its $z$-derivative $\varepsilon'_{33}$ are computed at the centre of the plate C. Then, according to the nature of the strain field, equation~\eqref{eq:normal_strain}, the derivative of the normal \WF{} is computed using:
\begin{equation}
  \varphi'_{33}(z) = \frac{\varepsilon_{33}(\tfrac{a}{2},\tfrac{b}{2},z)-\varepsilon_{33}(\tfrac{a}{2},\tfrac{b}{2},0^-)}{\varepsilon'_{33}(\tfrac{a}{2},\tfrac{b}{2},0^-)}
\end{equation}
The bending of various laminates has been investigated including those of this study, and the case where $\varepsilon'_{33}(\tfrac{a}{2},\tfrac{b}{2},0^-)=0$ has not yet occurred.
\par
As the normal \WF{} must verify $\varphi_{33}(0)=0$, it is computed using:
\begin{equation}
  \varphi_{33}(z) = \int_0^z \varphi'_{33}(\zeta)\text{d}\zeta
\end{equation}
\subsection{Computation of the $\varphi_{\alpha\beta}(z)$ warping functions}
\label{sec:InPlaneWF}
Considering equation~\eqref{eq:transverse_stresses2}, we see that the $\varphi'_{\alpha\beta}$ are directly linked to the $\sigma_{\alpha3}$. Introducing the transverse shear stresses $\sigma^0_{\delta 3}(x,y)$ at $z=0$ into this equation lead to
\begin{equation}
  \sigma_{\alpha3}(z) = C_{\alpha3\beta3}(z)\Big(4\varphi'_{\beta\gamma}(z) S_{\gamma3\delta3}(0) \sigma^0_{\delta 3}+z\varepsilon^0_{33,\beta}+\varphi_{33}(z)\kappa^0_{33,\beta}\Big)
\end{equation}
where $S_{\gamma3\delta3}$ are components of the compliance tensor. 
\par
This can be written
\begin{equation}\label{eq:psiprime}
  \sigma_{\alpha3}(z) = \Psi'_{\alpha\beta}(z)\sigma^0_{\beta 3} + C_{\alpha3\beta3}(z)\Big(z\varepsilon^0_{33,\beta}+\varphi_{33}(z)\kappa^0_{33,\beta}\Big)
\end{equation}
where:
\begin{equation}\label{eq:linkphipsi}
  \Psi'_{\alpha\beta}(z) =4 C_{\alpha3\delta3}(z) \varphi'_{\delta \gamma}(z) S_{\gamma3\beta3}(0)
\end{equation}
The $\Psi'_{\alpha\beta}(z)$ cannot be issued directly from equation~\eqref{eq:psiprime} because there are four functions to be determined from two stress variations, leading to infinitely many solutions. The main idea is to issue the four functions $\Psi'_{\alpha\beta}(z)$ from the transverse shear stresses in two separate locations on the plate, points A and B.
Since the deformation of the plate is of the form~\eqref{eq:BiSineDisplacementField}, the transverse shear stresses at the reference plane are of the form:
\begin{empheq}[left=\empheqlbrace]{align}
		\sigma^0_{13}(x,y) &= s_{13} \cos (\xi x) \sin (\eta y)  + \overline{s}_{13} \sin (\xi x) \cos (\eta y)  \\
		\sigma^0_{23}(x,y) &= s_{23} \sin (\xi x) \cos (\eta y)  + \overline{s}_{23} \cos (\xi x) \sin (\eta y) 
\end{empheq}
We can also write:
\begin{empheq}[left=\empheqlbrace]{align}
		\varepsilon^0_{33}(x,y) &= e_{33} \sin (\xi x) \sin (\eta y)  + \overline{e}_{33} \cos (\xi x) \cos (\eta y)  \\
		\kappa^0_{33}(x,y)      &= k_{33} \sin (\xi x) \sin (\eta y)  + \overline{k}_{33} \cos (\xi x) \cos (\eta y) 
\end{empheq}
These shear stresses are evaluated at points $A$ and $B$ for which:
\begin{itemize}
	\item[--] at point A, $x=a/2$ and $y=0$,  then $\sigma^0_{13}(A) = \overline{s}_{13}$, $\sigma^0_{23}(A) = s_{23}$,
            $\varepsilon^0_{33,1}=-\xi\overline{e}_{33}$, $\varepsilon^0_{33,2}=\eta e_{33}$,
            $\kappa^0_{33,1}=-\xi\overline{k}_{33}$ and $\kappa^0_{33,2}=\eta k_{33}$
	\item[--] at point B, $x=0$ and $y=b/2$, then $\sigma^0_{13}(B) = s_{13}$, $\sigma^0_{23}(B) = \overline{s}_{23}$
            $\varepsilon^0_{33,1}=\xi e_{33}$, $\varepsilon^0_{33,2}=-\eta\overline{e}_{33}$,
            $\kappa^0_{33,1}=\xi k_{33}$ and $\kappa^0_{33,2}=-\eta \overline{k}_{33}$
\end{itemize}
Setting these local values into formula~\eqref{eq:psiprime} leads to the following system:
\begin{gather}
\left[
\begin{matrix}
s_{13} & 0 & \overline{s}_{23} & 0 \\
0 & s_{23} & 0 & \overline{s}_{13} \\
\overline{s}_{13} & 0  & s_{23} & 0 \\
0 & \overline{s}_{23} & 0 &s_{13} \\
\end{matrix}
\right]
\left\{
\begin{matrix}
\Psi'_{11} \\
\Psi'_{22} \\
\Psi'_{12} \\
\Psi'_{21}
\end{matrix}
\right\}
=
\left\{
\begin{matrix}
\sigma_{13}(B) \\
\sigma_{23}(A) \\
\sigma_{13}(A) \\
\sigma_{23}(B)
\end{matrix}
\right\}
-
\nonumber\\
\left[
\begin{matrix}
\xi  C_{1313} & \eta C_{1323} & \xi  C_{1313} & \eta C_{1323}  \\
\eta C_{2323} & \xi  C_{2313} & \eta C_{2323} & \xi  C_{2313}  \\
\eta C_{1323} & \xi  C_{1313} & \eta C_{1323} & \xi  C_{1313}  \\
\xi  C_{2313} & \eta C_{2323} & \xi  C_{2313} & \eta C_{2323}  
\end{matrix}
\right]
\left\{
\begin{matrix}
   z e_{33} \\
  -z \overline{e}_{33}\\
   \varphi_{33}(z) k_{33} \\
  -\varphi_{33}(z) \overline{k}_{33}
\end{matrix}
\right\}
\end{gather}
The $\Psi'_{\alpha\beta}(z)$ are obtained from the resolution of this system; $\varphi'_{\alpha\beta}(z)$ are then obtained using the reciprocal of equation~\eqref{eq:linkphipsi}:
\begin{equation}
  \varphi'_{\alpha\beta}(z)=4 S_{\alpha3\delta3}(z) \Psi'_{\delta \gamma}(z) C_{\gamma3\beta3}(0)
\end{equation}
Then, integrating the $\varphi'_{\alpha\beta}(z)$ so that $\varphi_{\alpha\beta}(0)=0$ gives the four \WFs{} $\varphi_{\alpha\beta}(z)$.
\subsection{Practical considerations}\label{sec:Practical}
The way to use this theory in practical cases needs to be discussed. The \WFs{} always exist, once the materials, the lamination sequence, the length-to-thickness ratios $a/h$ and $b/h$, frequency $\omega$, and the wavenumbers $m$ and $n$ are given. The \WFs{} are computed using the above procedure, which is based on an analytical solution of a simply supported bending problem. A question which may be addressed is: how to link a general problem with the specific one from which will be issued the \WFs{}? While waiting for more relevant strategies, one has to choose $a/h$ and $b/h$ (and $\omega$, $m$ and $n$ if necessary) which are representative of the studied structure, regardless of the nature of applied boundary conditions and applied load, and then compute the \WFs{} with the above process. These \WFs{} implicitly depend on all previously enumerated parameters, which is not the case of most of the kinematics proposed until this date. In this paper, it is shown that the model which uses these \WFs{} performs better than other explicit kinematics of same order for the solving of simply supported problems. For more general problems, involving other boundary conditions, various loads, complex geometries, the results can depend on the choices mentioned above. Hence further studies need to be done to compare the different approaches.
\par
Remark: It is even possible not to determine the aforementioned parameters, and set default values, for example $a/h=b/h=100$, $\omega=0$, $m=n=1$. Doing this leads to usable \WFs{}, which are similar in shape to those of the two zig-zag theories which have been retained here for comparison. This default setting has not been studied.
\section{Solving method by a Navier-like procedure}\label{sec:Navier}
A Navier-like procedure is implemented to solve both static and dynamic problems for a simply supported plate. The dynamic study is restricted to the search of the first natural frequency. For laminates which are not of cross-ply nor anti-symmetrical angle-ply types, the simply supported boundary condition is replaced by a \emph{globally simply supported} condition. In this case, the plate could have a non-null deflection on its edges with respect of an antisymmetry with the opposite edge, and the first vibration mode splits into two modes, see reference~\cite{Loredo2014} for more details.
\par
The Fourier series is limited to one term, hence the generalized displacement field is
\begin{equation}
	\label{eq:BiSineDisplacementField}
	\left\{
	\begin{array}{c}
		u_1 \\
		u_2 \\
		w \\
		\gamma_{13} \\
		\gamma_{23} \\
		\varepsilon_{33} \\
		\kappa_{33}
	\end{array}
	\right\}
	=
  \left\{
	\begin{array}{clllll}
		u^{mn}_1             & \cos (\xi x) & \sin (\eta y)  &+	 \overline{u}^{mn}_1             & \sin (\xi x) & \cos (\eta y)  \\
		u^{mn}_2             & \sin (\xi x) & \cos (\eta y)  &+	 \overline{u}^{mn}_2             & \cos (\xi x) & \sin (\eta y)  \\
		w^{mn}               & \sin (\xi x) & \sin (\eta y)  &+	 \overline{w}^{mn}               & \cos (\xi x) & \cos (\eta y)  \\
		\gamma^{mn}_{13}     & \cos (\xi x) & \sin (\eta y)  &+	 \overline{\gamma}^{mn}_{13}     & \sin (\xi x) & \cos (\eta y)  \\
		\gamma^{mn}_{23}     & \sin (\xi x) & \cos (\eta y)  &+	 \overline{\gamma}^{mn}_{23}     & \cos (\xi x) & \sin (\eta y)  \\
		\varepsilon^{mn}_{33}& \sin (\xi x) & \sin (\eta y)  &+	 \overline{\varepsilon}^{mn}_{33}& \cos (\xi x) & \cos (\eta y)  \\
		\kappa^{mn}_{33}     & \sin (\xi x) & \sin (\eta y)  &+	 \overline{\kappa}^{mn}_{33}     & \cos (\xi x) & \cos (\eta y)  
  \end{array}
	\right\}
\end{equation}
with
\begin{equation}\nonumber
\xi = \frac{m \pi}{a} \text{ and } \eta = \frac{n \pi}{b}
\end{equation}
where $m$ and $n$ are wavenumbers, set to $1$ in this study. Considering formula~\eqref{eq:BiSineDisplacementField}, the motion equations of section~\ref{sec:LaminateEquationsOfMotion} give a stiffness and a mass matrix, respectively $\mymat{K}$ and $\mymat{M}$, related to the vector $\myvect{U}=\{u_1^{mn} ,u_2^{mn} \dots \overline{\varepsilon}^{mn}_{33}, \overline{\kappa}^{mn}_{33}\}$. The static case is treated solving the linear system $\mymat{K}\myvect{U}=\myvect{F}$, where $\myvect{F}$ is a force vector containing $-q^{mn}$, $-r^{mn}$ and $-s^{mn}$ for its third, sixth and seventh components. Solving the dynamic case consists in researching the generalized eigenvalues for matrices $\mymat{K}$ and $\mymat{M}$.
\section{Reference models retained for comparisons}
\label{sec:RefModels}
The model presented in section~\ref{sec:Formulation}, which uses five \WFs{} issued from transverse shear and normal stresses of analytical solutions, will be denoted \tdfivewf{}. The results obtained with the \tdfivewf{} model are compared to those obtained with different models issued from the literature and with the exact analytical solution. These reference models are presented below.
\subsection{Exact solution (Exa)}\label{sec:Exa}
Each studied case is solved by a state-space method described in reference~\cite{Loredo2014}. This method is a generalization for general lamination sequences of existing methods for cross-ply and antisymmetric angle-ply lamination sequences. 
In this work, the exact solution is used to obtain deflections, stresses and natural frequencies taken as reference for comparisons, but it is also used to create sets of \WFs{} for both \tdfourwf{} and \tdfivewf{} models, as explained in section~\ref{sec:WF}. The corresponding solution is denoted Exa in the following text and in tables.
\subsection{Models without normal deformation}\label{sec:ModelsWithoutND}
Some more or less classical models have been chosen for comparison matters. They offer the advantage to be easily simulated with the present model when appropriate sets of \WFs{} are selected:
\setlength{\leftmargini}{0cm}
\begin{description}
	\item[--]ToSDT, Third-order Shear Deformation Theory: often called Reddy's third order theory, verifies that transverse shear stresses are null at the top and bottom faces of the plate. It is simulated using the following \WFs{} ($\delta^K_{\alpha \beta}$ is the Kronecker's delta symbol):
	\begin{equation}
		\varphi_{\alpha \beta}(z) = \delta^K_{\alpha \beta} \left(z - \frac{4}{3}\frac{z^3}{h^2}\right)
	\end{equation}
	\item[--]\tozzfour{}, Third-order Zig-Zag model with 4 \WFs{}: This formulation, presented in references~\cite{Cho1993, Kim2007} consists in superimposing a cubic displacement field, which permits the transverse shear stresses to be null at the top and bottom faces of the laminate, to a zig-zag displacement field issued from the continuity of the transverse shear stresses at layer interfaces. This model can be rewritten with the following formulas. Let denote the $\ell$-th interface coordinate by $z=\zeta^{\ell}$, then the layer $\ell$ is situated between $z=\zeta^{\ell-1}$ and $z=\zeta^{\ell}$. The \WFs{} are the following piecewise-defined functions
\begin{equation}
  \left\{
  \begin{array}{l}
    \varphi^\ell_{\alpha\beta}(z) = a_{\alpha\beta}z^3+b_{\alpha\beta}z^2+c_{\alpha\beta}^\ell z + d_{\alpha\beta}^\ell \\
    \text{for }\zeta^{\ell-1} \le z \le \zeta^{\ell}
  \end{array}
  \right.
\end{equation}
the $8(N+1)$ constants are determined solving the $8(N+1)$ equations
  \begin{equation}\label{eq:constants}
    \left\{
    \begin{array}{l}
      \varphi^{\ell_0}_{\alpha\beta}(0)=0
      \, , \;
      \varphi'^{\ell_0}_{\alpha\beta}(0)=\delta_{\alpha\beta}
      \, , \;
      \varphi^\ell_{\alpha\beta}(\zeta^{\ell})=\varphi^{\ell+1}_{\alpha\beta}(\zeta^{\ell}) \\
      C^1_{\alpha3\beta3}\varphi'^1_{\beta\gamma}(-h/2)=0
      \, , \;
      C^N_{\alpha3\beta3}\varphi'^N_{\beta\gamma}(h/2)=0 \\
      C^\ell_{\alpha3\beta3}\varphi'^\ell_{\beta\gamma}(\zeta^{\ell})=C^{\ell+1}_{\alpha3\beta3}\varphi'^{\ell+1}_{\beta\gamma}(\zeta^{\ell})
    \end{array}
    \right.
  \end{equation}
where $\ell_0$ is the lower layer containing the $z=0$ plane.
	\item[--]\sizzfour{}, Sine Zig-Zag model with 4 \WFs{}: This model, inspired from the Beakou-Touratier model~\cite{Beakou1993}, verifies the continuity of transverse shear stresses at the layers' interfaces. The original model has been enhanced in this study in order to obtain a good behaviour for general lamination sequences, both sine and cosine functions being now present in the $4$ \WFs{}. With the same definitions for layer coordinates than in the previous model, the \WFs{} are the following piecewise-defined functions:
\begin{equation}
  \left\{
  \begin{array}{l}
    \varphi^\ell_{\alpha\beta}(z) = a_{\alpha\beta}\sin\left(\frac{\pi z}{h}\right)+b_{\alpha\beta}\cos\left(\frac{\pi z}{h}\right)+c_{\alpha\beta}^\ell z + d_{\alpha\beta}^\ell \\
    \text{for }\zeta^{\ell-1} \le z \le \zeta^{\ell}
  \end{array}
  \right.
\end{equation}
The constants are determined with the method already described (see formulas~\eqref{eq:constants}).
\item[--]\tdfourwf{}, this model, which has been briefly presented in the introduction, is issued from reference~\cite{Loredo2014b}.

\end{description}
\subsection{Models with normal deformation}\label{sec:ModelsWithND}
The three first models described in the precedent section, have been extended in order to take into account a normal deformation. This has been done considering a modified displacement field including equation~\eqref{eq:normal_depl}, with the following choice of $\varphi_{33}(z)$:
	\begin{equation}\label{eq:phi33}
		\varphi_{33}(z) = \frac{z^2}{2}
	\end{equation}
These three models will be denoted EToSDT, \tozzfive{} and \sizzfive{}.%
\section{Numerical results}
\label{sec:nresults}
This section proposes the study of five laminate configurations including an isotropic plate, two single layer orthotropic plates and a sandwich panel. Three materials are involved, an isotropic material for the isotropic plate, an orthotropic composite material for all laminates and an honeycomb-type material for the core of the sandwich panel. All the properties are given in table~\ref{tab:matprop}. For all computations, the loading is divided into two equal parts which are applied to the top and bottom faces.
\par
\begin{table}[tb]
  \centering
  \footnotesize
  \setlength{\tabcolsep}{2pt}
  \begin{tabular}{lcccccccccc}
                       &   $E_1$   &    $E_2$   &    $E_3$    &  $G_{23}$  &  $G_{13}$  &  $G_{12}$  & $\nu_{23}$ & $\nu_{13}$ & $\nu_{12}$ &  $\rho$  \\	\hline
    Isotropic mat. (i) & $E^i$     &    $E^i$   &    $E^i$    &  $0.4E^i$  &  $0.4E^i$  & $0.4E^i$   &   $0.25$   &    $0.25$  &   $0.25$   & $\rho^i$ \\
    Composite (c)      & $25E^c_2$ &   $E^c_2$  &   $E^c_2$   & $0.2E^c_2$ & $0.5E^c_2$ & $0.5E^c_2$ &   $0.25$   &   $0.25$   &   $0.25$   & $\rho^c$ \\
    Honeycomb (h)      &  $E^h_2$  & $E^c_2/25$ & $12.5E^h_2$ & $1.5E^h_2$ & $1.5E^h_2$ & $0.4E^h_2$ &   $0.02$   &   $0.02$   &   $0.25$   &  $\rho^c/15$
	\end{tabular}
	\caption{Material properties.}
	 \label{tab:matprop}
\end{table}
\par
Deflections $w$, first natural frequencies $\omega$ and stresses $\sigma_{i3}$ are nondimensionalized using the following formulas
  \begin{equation}
    w^{*}=100\frac{E_2^{\text{ref}} h^3}{(-q)a^4}w 
    \,\text{,}\enspace
    \omega^{*}=\frac{a^2}{h}\sqrt{\frac{\rho^{\text{ref}}}{E_2^{\text{ref}}}}\omega
    \,\text{,}\enspace
    \sigma_{i3}^{*}=10\frac{h}{(-q)a}\sigma_{i3}
	\label{eq:nondimd}
\end{equation}
where $E_2^{\text{ref}}$ and $\rho^{\text{ref}}$ are taken as values of the core material for the sandwich and as values of the corresponding material for other cases.
\par
All transverse and normal stresses appearing in tables and figures have been computed integrating equilibrium equations, in accordance with the in-plane kinematics of each model.
\subsection{Square isotropic plate}\label{sec:iso}
For this first study, the \WFs{} of different models do not strongly differ, it is the reason why they are not plotted. Such comparisons are let for the following examples. It can be seen in table~\ref{tab:iso} that for $a/h=2$, all ``extended'' models give better results for the deflection than the original model they are issued from. The results on transverse stresses are less good but quite comparable. Replacing the Poisson's coefficient value of 0.25 by 0.35 leads to a change in the ranking of the ZZ models. Table~\ref{tab:iso} also shows that results of models without normal deformation are better if $a/h$ takes higher values.
%
%
%
%
%
\begin{table}[tb]%
	\centering
	\footnotesize
  \setlength{\tabcolsep}{3pt}
	\begin{tabular}{c@{~}l|c@{~}c|l@{~}c|c@{~}c|c@{~}c}
    a/h & Model & $w^{*}$ & \% & $\sigma_{13}(B)$ & \% & $\sigma_{23}(A)$ & \% & $\omega^{*}$ & \% \\ 

    \hline
    2 & ToSDT & $6.6282$ & $+9.24$ & $2.2585$ & $-0.81$ & $2.2585$ & $-0.81$ & $3.7249$ & $-0.86$ \\ 
  & \tozzfour{} & $6.6282$ & $+9.24$ & $2.2585$ & $-0.81$ & $2.2585$ & $-0.81$ & $3.7249$ & $-0.86$ \\ 
  & \sizzfour{} & $6.6096$ & $+8.93$ & $2.2421$ & $-1.52$ & $2.2421$ & $-1.52$ & $3.7288$ & $-0.76$ \\ 
  & \tdfourwf{} & $6.6347$ & $+9.35$ & $2.2756$ & $-0.05$ & $2.2756$ & $-0.05$ & $3.7239$ & $-0.89$ \\ 
  & EToSDT & $6.0989$ & $+0.52$ & $2.2448$ & $-1.41$ & $2.2448$ & $-1.41$ & $3.7712$ & $+0.37$ \\ 
  & \tozzfive{} & $6.0989$ & $+0.52$ & $2.2448$ & $-1.41$ & $2.2448$ & $-1.41$ & $3.7712$ & $+0.37$ \\ 
  & \sizzfive{} & $6.0724$ & $+0.08$ & $2.2267$ & $-2.20$ & $2.2267$ & $-2.20$ & $3.7793$ & $+0.59$ \\ 
  & \tdfivewf{} & $6.0674$ & $-0.00$ & $2.2769$ & $+0.00$ & $2.2769$ & $+0.00$ & $3.7573$ & $+0.00$ \\ 
  & Exact & $6.0675$ &  & $2.2769$ &  & $2.2769$ &  & $3.7572$ &  \\ 

    \hline
    4 & ToSDT &       $3.8335$ & $+2.73$ & $2.3547$ & $-0.30$ & $2.3547$ & $-0.30$ & $4.9620$ & $-0.45$ \\ 
  & \tozzfour{} & $3.8335$ & $+2.73$ & $2.3547$ & $-0.30$ & $2.3547$ & $-0.30$ & $4.9620$ & $-0.45$ \\ 
  & \sizzfour{} & $3.8313$ & $+2.67$ & $2.3505$ & $-0.48$ & $2.3505$ & $-0.48$ & $4.9633$ & $-0.43$ \\ 
  & \tdfourwf{} & $3.8336$ & $+2.73$ & $2.3557$ & $-0.26$ & $2.3557$ & $-0.26$ & $4.9620$ & $-0.45$ \\ 
  & EToSDT      & $3.7296$ & $-0.06$ & $2.3500$ & $-0.50$ & $2.3500$ & $-0.50$ & $4.9890$ & $+0.09$ \\ 
  & \tozzfive{} & $3.7296$ & $-0.06$ & $2.3500$ & $-0.50$ & $2.3500$ & $-0.50$ & $4.9890$ & $+0.09$ \\ 
  & \sizzfive{} & $3.7238$ & $-0.21$ & $2.3452$ & $-0.71$ & $2.3452$ & $-0.71$ & $4.9927$ & $+0.16$ \\ 
  & \tdfivewf{} & $3.7317$ & $-0.00$ & $2.3619$ & $+0.00$ & $2.3619$ & $+0.00$ & $4.9846$ & $+0.00$ \\ 
  & Exact & $3.7317$ &  & $2.3619$ &  & $2.3619$ &  & $4.9846$ &  \\ 

    \hline
    10 & ToSDT       & $3.0392$ & $+0.48$ & $2.3821$ & $-0.05$ & $2.3821$ & $-0.05$ & $5.6940$ & $-0.10$ \\ 
   & \tozzfour{} & $3.0392$ & $+0.48$ & $2.3821$ & $-0.05$ & $2.3821$ & $-0.05$ & $5.6940$ & $-0.10$ \\ 
   & \sizzfour{} & $3.0390$ & $+0.47$ & $2.3814$ & $-0.08$ & $2.3814$ & $-0.08$ & $5.6942$ & $-0.10$ \\ 
   & \tdfourwf{} & $3.0392$ & $+0.48$ & $2.3821$ & $-0.05$ & $2.3821$ & $-0.05$ & $5.6940$ & $-0.10$ \\ 
   & EToSDT      & $3.0239$ & $-0.02$ & $2.3813$ & $-0.09$ & $2.3813$ & $-0.09$ & $5.7005$ & $+0.01$ \\ 
   & \tozzfive{} & $3.0239$ & $-0.02$ & $2.3813$ & $-0.09$ & $2.3813$ & $-0.09$ & $5.7005$ & $+0.01$ \\ 
   & \sizzfive{} & $3.0231$ & $-0.05$ & $2.3805$ & $-0.12$ & $2.3805$ & $-0.12$ & $5.7013$ & $+0.03$ \\ 
   & \tdfivewf{} & $3.0246$ & $-0.00$ & $2.3834$ & $+0.00$ & $2.3834$ & $+0.00$ & $5.6998$ & $+0.00$ \\ 
   & Exact & $3.0246$ &  & $2.3834$ &  & $2.3834$ &  & $5.6998$ &  \\ 

	\end{tabular}
	\normalsize
	\caption{Comparison between the different models for the square $[iso]$ isotropic plate with various length-to-thickness ratios.}
	\label{tab:iso}
\end{table}
\subsection{Square $[0]$ composite plate}\label{sec:p0}
In table~\ref{tab:p0}, results show an inverse tendency than for the previous case: all ``extended'' models show less good results for the deflection than original models, except the \tdfivewf{} model. All models, except the \tdfourwf{} and \tdfivewf{} models which have material-sensitive \WFs{}, have the same \WFs{} for an orthotropic single layer plate than for an isotropic single layer one. It can be seen that small differences in the kinematic assumptions of models can have great influence in results. The \tdfourwf{} and \tdfivewf{} models have different \WFs{} for the $x$ and $y$ directions, $\phi_{11}(z)\ne\phi_{22}(z)$, as it can be seen in figure~\ref{fig:p0_phi}. This is due to different shear/longitudinal modulus ratios in the $x$ and $y$ directions. For all the other models $\varphi_{11}(z)=\varphi_{22}(z)$. The \WFs{} of classical models are not presented because they are the same than those of the corresponding extended models. Those of the \tdfourwf{} have not been presented for clarity. Further, even if the differences on the \WFs{} are small, strong differences can be observed when stresses are computed, as shown in figure~\ref{fig:p0_sigma}. Due to their formulation, the \tozzfive{} and the EToSDT coincide for a one layer plate, and the \sizzfive{} model do not strongly differ from the two previous. These three models give transverse shear stresses that differ from the exact solution, especially in the $x$ direction. This shows that the $z-4z^3/(3h^2)$ function, and also the sine function of model \sizzfive{}, are not able to fit the behaviour of material with a shear/longitudinal modulus ratio of $0.02$ that differs strongly from the isotropic case. On the contrary, as $G_{23}/E_2=0.2$, value closer to the previous isotropic ratio of $0.4$, the $\sigma_{22}$ stress is better fitted by these models. This may also explain why, unlike the previous case, the ``extended'' models give worse deflection values than the corresponding originals ones. The \tdfivewf{} model, with its material sensitive formulation, predicts the good values for transverse stresses.
\iftoggle{submission}{}{\tikzsetnextfilename{p0_phi}}
\begin{SmartFigure}[tb]
\setlength{\abovecaptionskip}{0pt plus 0pt minus 0pt}
  \centering
  \iftoggle{submission}{
    \includegraphics{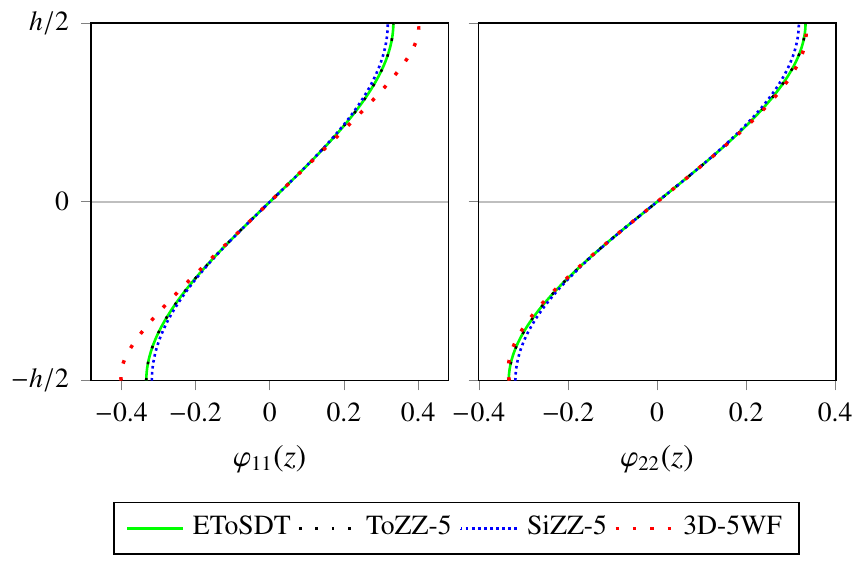}
  }{
    \input{Figures/p0_phi.tex}
  }
  \caption{Transverse shear \WFs{} of the $[0]$ single ply composite plate with $a/h=2$ for each model.}%
  \label{fig:p0_phi}
\end{SmartFigure}
\iftoggle{submission}{}{\tikzsetnextfilename{p0_sigma}}
\begin{SmartFigure}[tb]
  \centering
  \iftoggle{submission}{
    \includegraphics{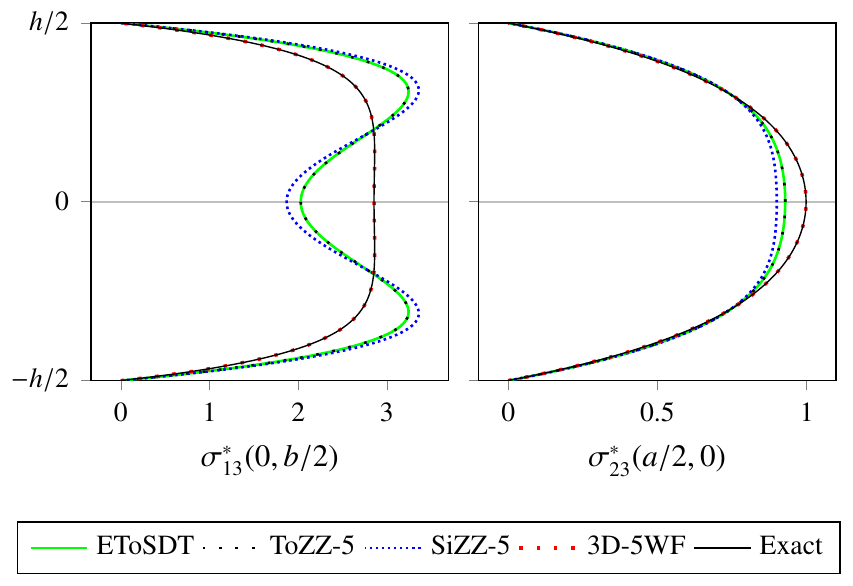}
  }{
    \input{Figures/p0_sigma.tex}
  }
  \caption{Nondimensionalized transverse shear stresses of the $[0]$ single ply composite plate with $a/h=2$ for each model.}%
  \label{fig:p0_sigma}
\end{SmartFigure}
\iftoggle{submission}{}{\tikzsetnextfilename{p0_33}}
\begin{SmartFigure}[tb]
  \centering
  \iftoggle{submission}{
    \includegraphics{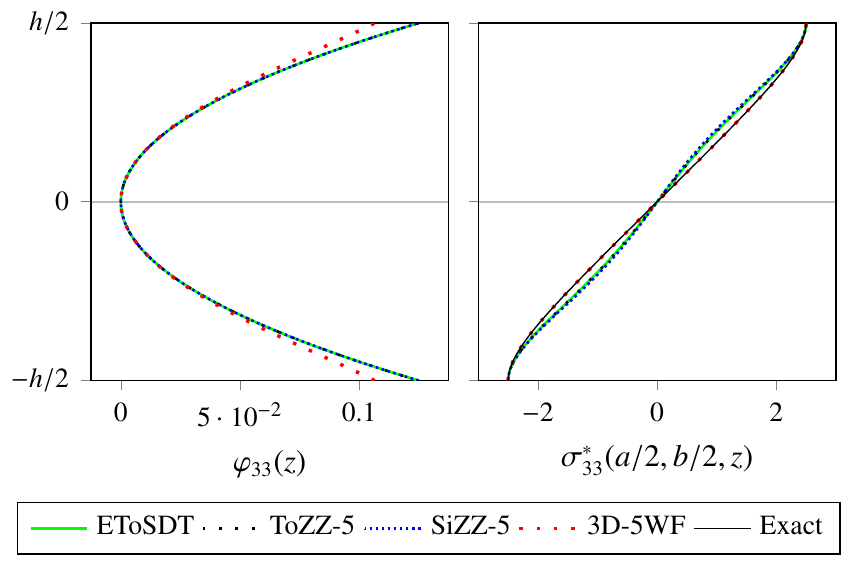}
  }{
    \input{Figures/p0_33.tex}
  }
  \caption{Normal WF and nondimensionalized normal stress for the $[0]$ single ply composite plate with $a/h=2$ for each model. By definition, the exact $\varphi_{33}$ coincides with the \tdfivewf{} one, hence it is not plotted.}%
  \label{fig:p0_33}
\end{SmartFigure}
\begin{table}[tb]%
	\centering
	\footnotesize
  \setlength{\tabcolsep}{3pt}
	\begin{tabular}{c@{~}l|c@{~}c|l@{~}c|c@{~}c|c@{~}c}
    a/h & Model & $w^{*}$ & \% & $\sigma_{13}(B)$ & \% & $\sigma_{23}(A)$ & \% & $\omega^{*}$ & \% \\ 

    \hline
    2 & ToSDT       & $4.5262$ & $+1.19$ & $2.0291$ & $-28.93$ & $0.85739$ & $-14.09$ & $4.6223$ & $+2.50$ \\ 
  & \tozzfour{} & $4.5262$ & $+1.19$ & $2.0291$ & $-28.93$ & $0.85739$ & $-14.09$ & $4.6223$ & $+2.50$ \\ 
  & \sizzfour{} & $4.4160$ & $-1.27$ & $1.8654$ & $-34.66$ & $0.82793$ & $-17.04$ & $4.6769$ & $+3.72$ \\ 
  & \tdfourwf{} & $4.7804$ & $+6.87$ & $3.1758$ & $+11.24$ & $0.95157$ & $-4.65$ & $4.5040$ & $-0.12$ \\ 
  & EToSDT      & $4.2714$ & $-4.51$ & $2.0287$ & $-28.94$ & $0.92807$ & $-7.00$ & $4.6396$ & $+2.89$ \\ 
  & \tozzfive{} & $4.2714$ & $-4.51$ & $2.0287$ & $-28.94$ & $0.92807$ & $-7.00$ & $4.6396$ & $+2.89$ \\ 
  & \sizzfive{} & $4.1689$ & $-6.80$ & $1.8691$ & $-34.53$ & $0.89980$ & $-9.84$ & $4.6949$ & $+4.12$ \\ 
  & \tdfivewf{} & $4.4727$ & $-0.01$ & $2.8556$ & $+0.03$ & $0.99800$ & $+0.00$ & $4.5095$ & $+0.00$ \\ 
  & Exact & $4.4730$ &  & $2.8549$ &  & $0.99796$ &  & $4.5093$ &  \\ 

	\end{tabular}
	\normalsize
	\caption{Comparison between the different models for the square $[0]$ composite plate with $a/h=2$.}
	\label{tab:p0}
\end{table}
\subsection{Square $[5]$ composite plate}\label{sec:p5}
When used to study cross-ply multilayered plates, the \tozzfive{} and \sizzfive{} zig-zag models have non null $\varphi_{12}$ and $\varphi_{21}$ functions. However, for an angle-ply single layer plate, these zig-zag models do not lead to coupling between the $x$ and $y$ directions, which can be seen as a limitation. Indeed, as we can see in figure~\ref{fig:p5_phi}, only the \tdfivewf{} model have non null $\varphi_{12}$ and $\varphi_{21}$ functions. It is probably the reason why the \tozzfive{} and \sizzfive{} models give poor estimates of $\sigma_{23}(B)$ and $\sigma_{13}(A)$, as can be seen in figure~\ref{fig:p5_sigma} and in table~\ref{tab:p5}.
\iftoggle{submission}{}{\tikzsetnextfilename{p5_phi}}
\begin{SmartFigure}[tb]
  \centering
  \iftoggle{submission}{
    \includegraphics{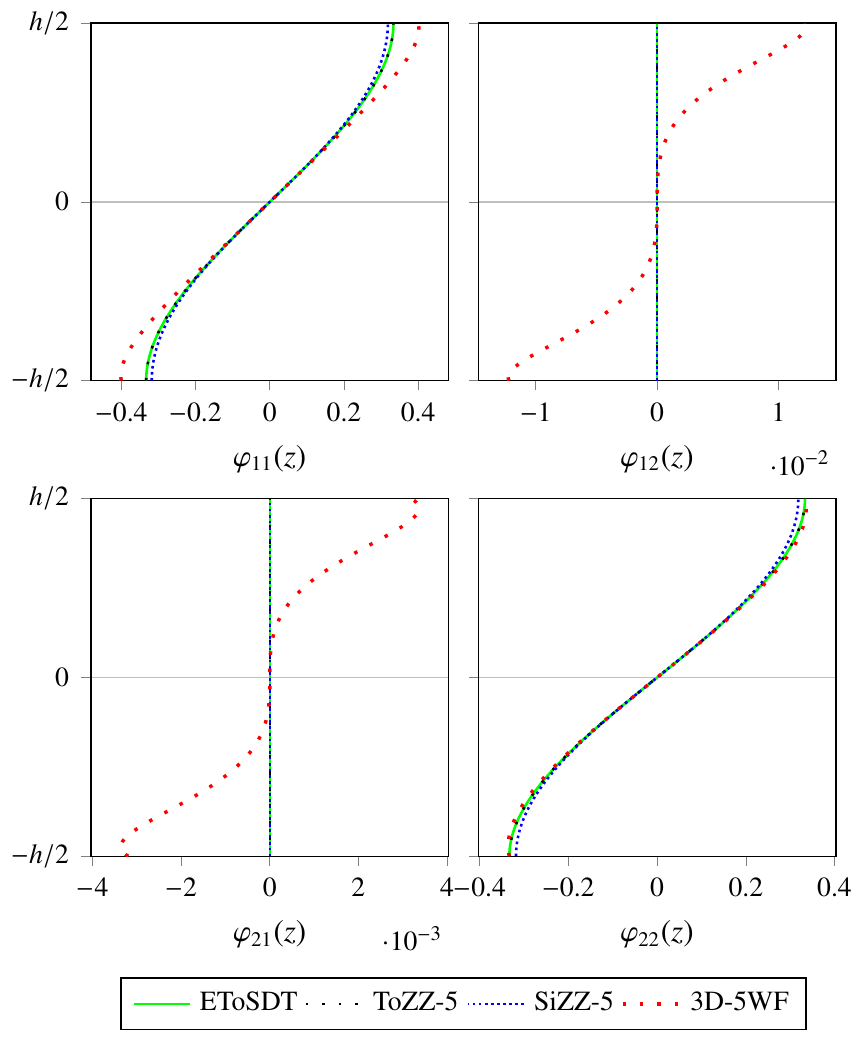}
  }{
    \input{Figures/p5_phi.tex}
  }
  \caption{Transverse shear \WFs{} of the $[5]$ single ply composite plate with $a/h=2$ for each model.}%
  \label{fig:p5_phi}
\end{SmartFigure}
\iftoggle{submission}{}{\tikzsetnextfilename{p5_sigma}}
\begin{SmartFigure}[tb]
  \centering
  \iftoggle{submission}{
    \includegraphics{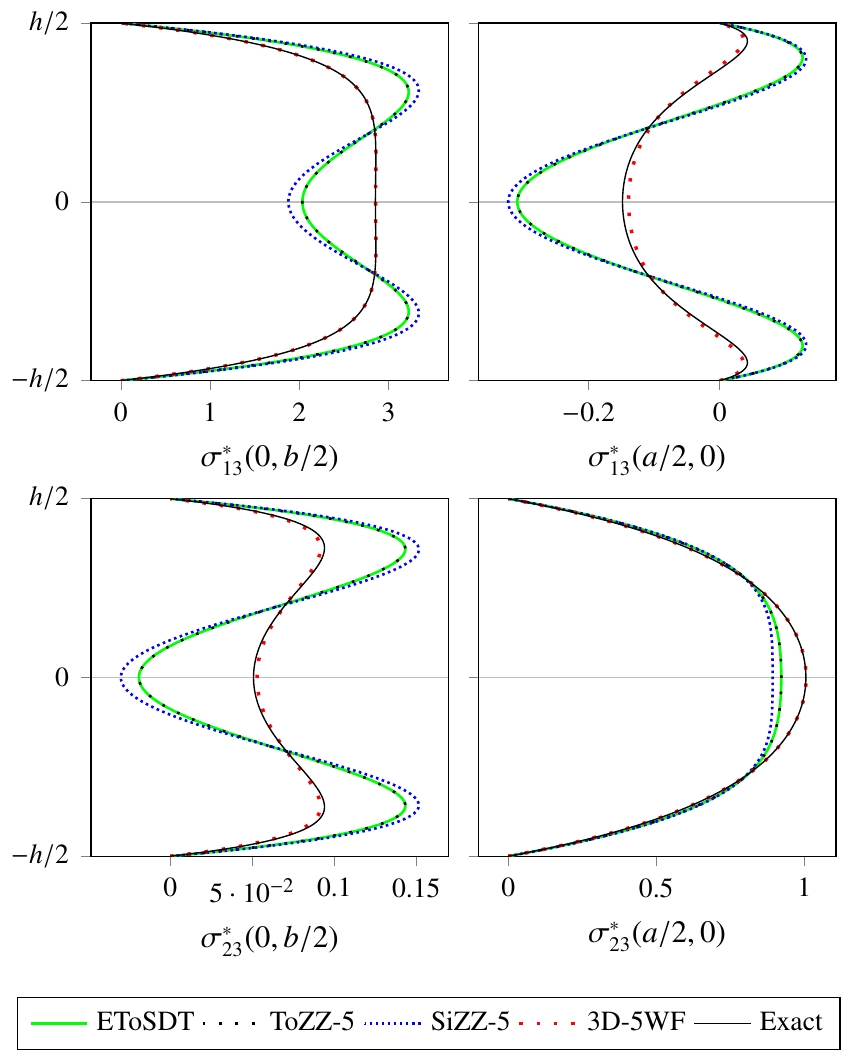}
  }{
    \input{Figures/p5_sigma.tex}
  }
  \caption{Nondimensionalized transverse shear stresses of the $[5]$ single ply composite plate with $a/h=2$ for each model.}%
  \label{fig:p5_sigma}
\end{SmartFigure}
\begin{table*}[tb]%
	\centering
	\footnotesize
	\begin{tabular}{@{~~}r@{~~}l|l@{~~}r|l@{~~}r|l@{~~}r|l@{~~}r|l@{~~}r|l@{~~}r@{~~}}
    a/h & Model & $w^{*}$ & \% & $\sigma_{13}(B)$ & \% & $\sigma_{23}(A)$ & \% & $\sigma_{23}(B)$ & \% & $\sigma_{13}(A)$ & \% & $\omega^{*}$ & \% \\ 

    \hline
    2 & ToSDT       & $4.5790$ & $+1.28$ & $2.0341$ & $-28.67$ & $0.85133$ & $-15.31$ & $-0.023725$ & $-146.74$ & $-0.32813$ & $+120.90$ & $4.3340$ & $+1.69$ \\ 
  & \tozzfour{} & $4.5790$ & $+1.28$ & $2.0341$ & $-28.67$ & $0.85133$ & $-15.31$ & $-0.023725$ & $-146.74$ & $-0.32813$ & $+120.90$ & $4.3340$ & $+1.69$ \\ 
  & \sizzfour{} & $4.4701$ & $-1.13$ & $1.8732$ & $-34.32$ & $0.82083$ & $-18.35$ & $-0.035293$ & $-169.54$ & $-0.34270$ & $+130.70$ & $4.3751$ & $+2.65$ \\ 
  & \tdfourwf{} & $4.8302$ & $+6.83$ & $3.1726$ & $+11.25$ & $0.96488$ & $-4.02$ & $0.079416$ & $+56.47$ & $-0.096157$ & $-35.27$ & $4.2546$ & $-0.17$ \\ 
  & EToSDT      & $4.3224$ & $-4.40$ & $2.0338$ & $-28.69$ & $0.92225$ & $-8.26$ & $-0.019114$ & $-137.66$ & $-0.30920$ & $+108.15$ & $4.3533$ & $+2.14$ \\ 
  & \tozzfive{} & $4.3224$ & $-4.40$ & $2.0338$ & $-28.69$ & $0.92225$ & $-8.26$ & $-0.019114$ & $-137.66$ & $-0.30920$ & $+108.15$ & $4.3533$ & $+2.14$ \\ 
  & \sizzfive{} & $4.2211$ & $-6.64$ & $1.8767$ & $-34.19$ & $0.89293$ & $-11.18$ & $-0.030121$ & $-159.35$ & $-0.32308$ & $+117.50$ & $4.3956$ & $+3.14$ \\ 
  & \tdfivewf{} & $4.5208$ & $-0.01$ & $2.8537$ & $+0.06$ & $1.0061$ & $+0.08$ & $0.052901$ & $+4.23$ & $-0.13928$ & $-6.24$ & $4.2672$ & $+0.12$ \\ 
  & Exact & $4.5212$ &  & $2.8519$ &  & $1.0053$ &  & $0.050756$ &  & $-0.14854$ &  & $4.2620$ &  \\ 

	\end{tabular}
	\normalsize
	\caption{Comparison between the different models for the square $[5]$ composite plate with $a/h=2$.}
	\label{tab:p5}
\end{table*}
\subsection{Square $[0/c/0]$ sandwich plate}\label{sec:Sandwich}
Structures exhibiting a high variation of stiffness through the thickness are pertinent benchmarks for plate theories. Consider a square sandwich plate with ply thicknesses $h_1=h_3=0.1 h$ and $h_2 = 0.8 h$. The face sheets are made of one ply of unidirectional composite and the core is constituted of a honeycomb-type material. Material properties are presented in table~\ref{tab:matprop}. Results presented in table~\ref{tab:Sandwich} show this time that the zig-zag models give correct values.
\par
\begin{figure}[tb]%
  \centering
  \iftoggle{submission}{
    \includegraphics{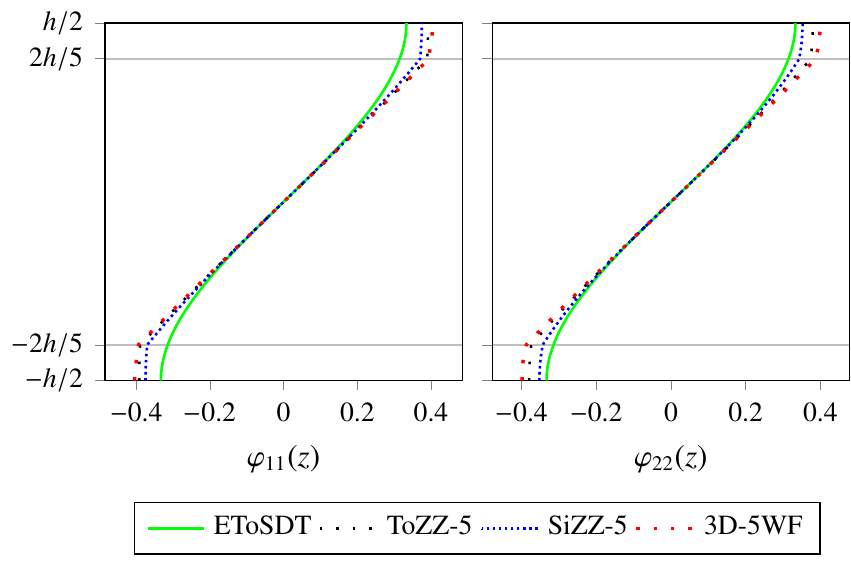}
  }{
    \input{Figures/Sandwich_phi.tex}%
  }
  \caption{Transverse shear \WFs{} of the $[0/c/0]$ sandwich plate with $a/h=2$ for each model.}%
  \label{fig:Sandwich_phi}
\end{figure}
\begin{figure}[tb]%
  \centering
  \iftoggle{submission}{
    \includegraphics{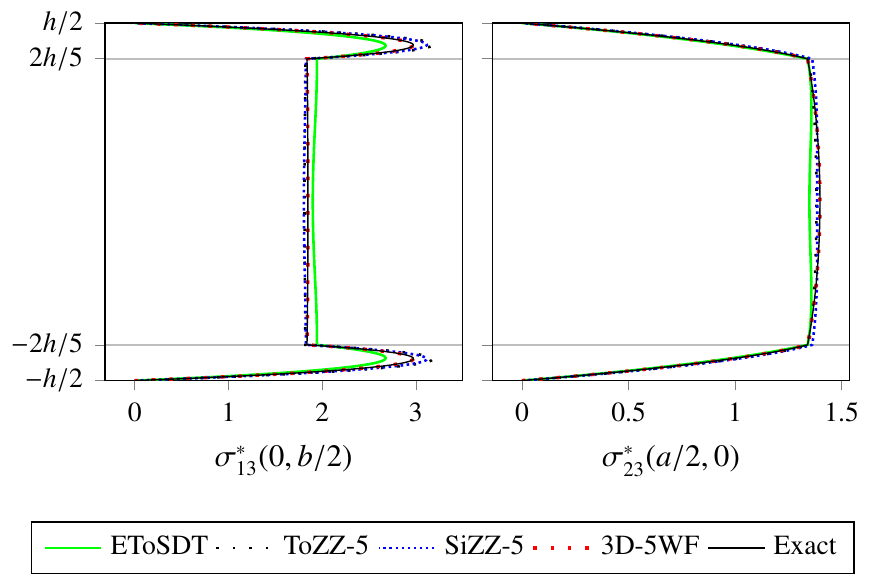}
  }{
    \input{Figures/Sandwich_sigma.tex}%
  }
  \caption{Nondimensionalized transverse shear stresses of the $[0/c/0]$ sandwich plate with $a/h=2$ for each model.}%
  \label{fig:Sandwich_sigma}
\end{figure}
\begin{figure}[tb]%
  \centering
  \iftoggle{submission}{
    \includegraphics{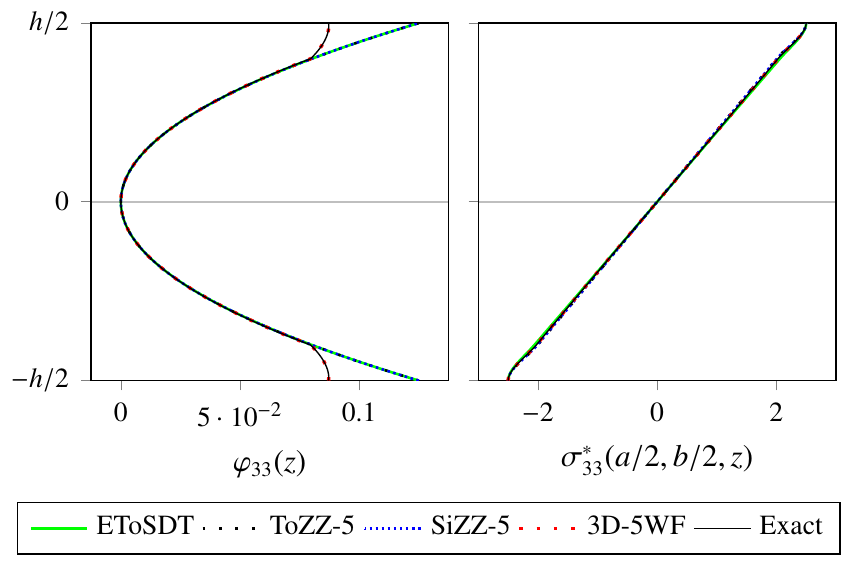}
  }{
    \input{Figures/Sandwich_33.tex}%
  }
  \caption{Normal WF and nondimensionalized normal stress for the $[0/c/0]$ sandwich plate with $a/h=2$ for each model. By definition, the exact $\varphi_{33}$ coincides with the \tdfivewf{} one, hence it is not plotted.}%
  \label{fig:Sandwich_33}
\end{figure}

Figure~\ref{fig:Sandwich_phi} shows the corresponding \WFs{} for $a/h=2$, for all plate models. Figure~\ref{fig:Sandwich_sigma} presents the transverse shear stresses at points A and B obtained for all models by the integration of equilibrium equations, compared to the exact solution, in the $a/h=2$ case. One can see that the $\sigma_{13}$ stress is quite overestimated in the skins by the zig-zag models. The Figure~\ref{fig:Sandwich_33} shows the $\varphi_{33}(z)$ \WF{} and the normal stress $\sigma_{33}(z)$ at point C. Although the $\varphi_{33}$ function of the \tdfivewf{} model differs from the $z^2/2$ function of formula~\eqref{eq:phi33}, only little differences can be seen on $\sigma_{33}$, all models giving correct estimates for the normal stress, in this case.
\iftoggle{submission}{}{\tikzsetnextfilename{Sandwich_phi}}
\iftoggle{submission}{}{\tikzsetnextfilename{Sandwich_sigma}}
\iftoggle{submission}{}{\tikzsetnextfilename{Sandwich_33}}
\begin{table}[tb]%
	\centering
	\footnotesize
  \setlength{\tabcolsep}{3pt}
	\begin{tabular}{c@{~}l|c@{~}c|l@{~}c|c@{~}c|c@{~}c}
    a/h & Model & $w^{*}$ & \% & $\sigma_{13}(B)$ & \% & $\sigma_{23}(A)$ & \% & $\omega^{*}$ & \% \\ 

    \hline
    2 & ToSDT & $0.85344$ & $-3.47$ & $1.9150$ & $+3.62$ & $1.3300$ & $-4.91$ & $5.4307$ & $+3.45$ \\ 
  & \tozzfour{} & $0.88896$ & $+0.55$ & $1.8320$ & $-0.87$ & $1.3570$ & $-2.97$ & $5.3281$ & $+1.50$ \\ 
  & \sizzfour{} & $0.89196$ & $+0.89$ & $1.8243$ & $-1.28$ & $1.3615$ & $-2.65$ & $5.3182$ & $+1.31$ \\ 
  & \tdfourwf{} & $0.91009$ & $+2.94$ & $1.8646$ & $+0.90$ & $1.3849$ & $-0.98$ & $5.2703$ & $+0.40$ \\ 
  & EToSDT & $0.83383$ & $-5.69$ & $1.8996$ & $+2.80$ & $1.3491$ & $-3.54$ & $5.4540$ & $+3.90$ \\ 
  & \tozzfive{} & $0.86928$ & $-1.68$ & $1.8115$ & $-1.97$ & $1.3824$ & $-1.16$ & $5.3451$ & $+1.82$ \\ 
  & \sizzfive{} & $0.87206$ & $-1.36$ & $1.8048$ & $-2.34$ & $1.3858$ & $-0.92$ & $5.3366$ & $+1.66$ \\ 
  & \tdfivewf{} & $0.88411$ & $-0.00$ & $1.8480$ & $+0.00$ & $1.3986$ & $+0.00$ & $5.2493$ & $+0.00$ \\ 
  & Exact & $0.88412$ &  & $1.8480$ &  & $1.3986$ &  & $5.2493$ &  \\ 

	\end{tabular}
	\normalsize
	\caption{Comparison between the different models for the square $[0/c/0]$ sandwich plate with a varying length to thickness ratio.}
	\label{tab:Sandwich}
\end{table}
\subsection{Square $[-45/0/45/90]_s$ composite plate}\label{sec:aero}
This example is given to test the behaviour of all models for a laminate with more than three layers. The \WFs{} are plotted in figure~\ref{fig:m45p0p45p90s_phi}. Except for the EToSDT model, which do not contains multilayer information, all the models present similar \WFs{}, included the $\varphi_{12}(z)$ and $\varphi_{21}(z)$ ones. However, despite the shapes of transverse stresses shown in figure~\ref{fig:m45p0p45p90s_sigma} which are similar, we can see in table~\ref{tab:m45p0p45p90s} that, due to its small relative value, $\sigma_{23}(B)$ can be poorly estimated. Nevertheless, considering the very low length-to-thickness ratio, results are not so bad for this example.
\iftoggle{submission}{}{\tikzsetnextfilename{m45p0p45p90s_phi}}
\begin{SmartFigure}[tb]
  \centering
  \iftoggle{submission}{
    \includegraphics{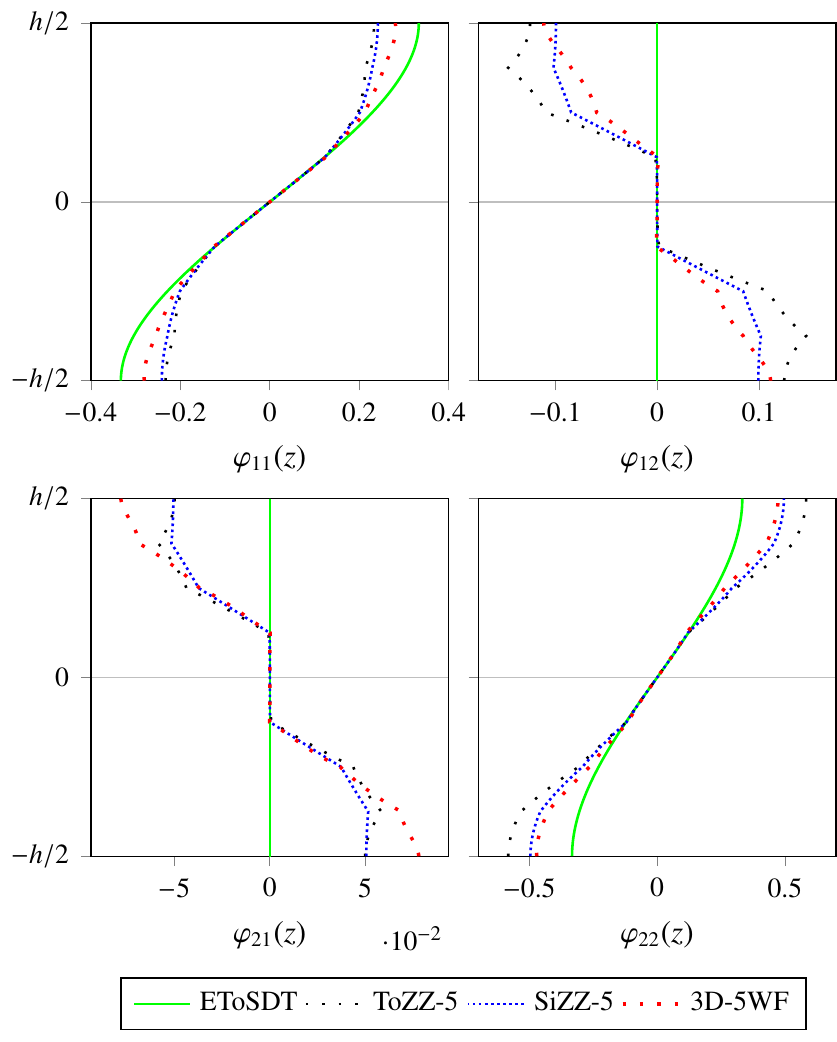}
  }{
    \input{Figures/m45p0p45p90s_phi.tex}
  }
  \caption{Transverse shear \WFs{} of the $[-45/0/45/90]_s$ composite plate with $a/h=2$ for each model.}%
  \label{fig:m45p0p45p90s_phi}
\end{SmartFigure}
\iftoggle{submission}{}{\tikzsetnextfilename{m45p0p45p90s_sigma}}
\begin{SmartFigure}[tb]
  \centering
  \iftoggle{submission}{
    \includegraphics{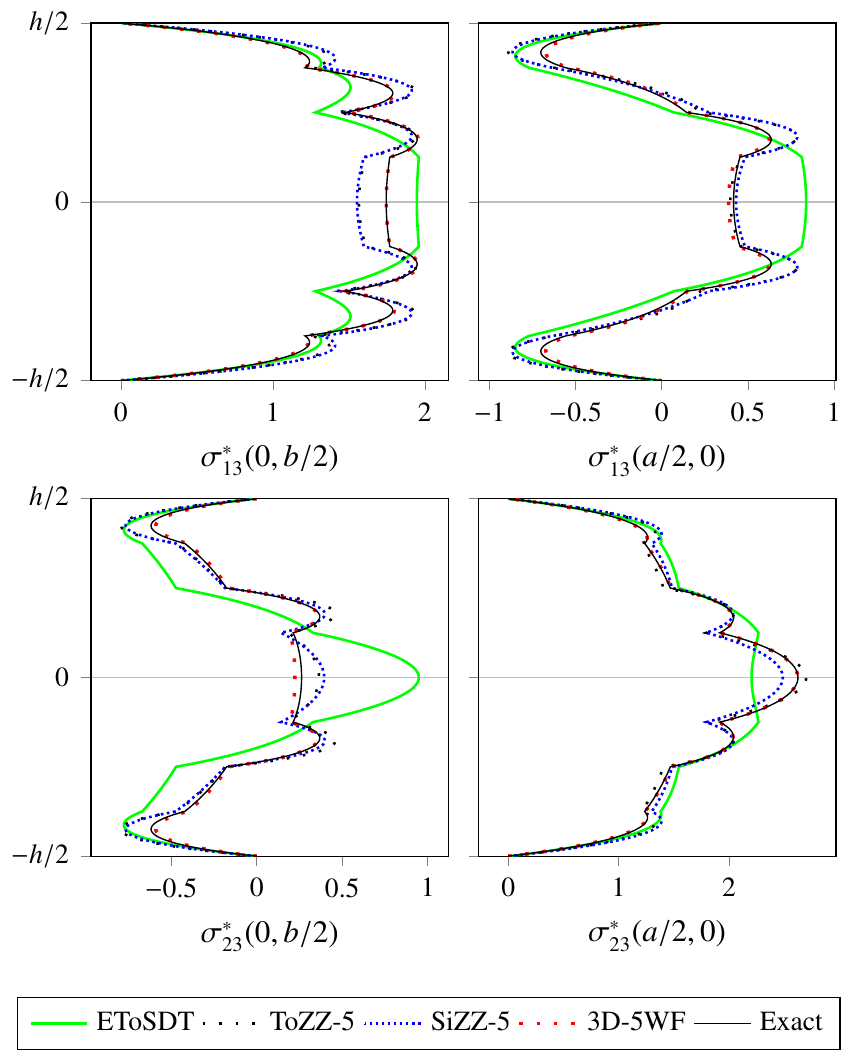}
  }{
    \input{Figures/m45p0p45p90s_sigma.tex}
  }
  \caption{Nondimensionalized transverse shear stresses of the $[-45/0/45/90]_s$ composite plate with $a/h=2$ for each model.}%
  \label{fig:m45p0p45p90s_sigma}
\end{SmartFigure}
\begin{table*}[tb]%
	\centering
	\footnotesize
	\begin{tabular}{@{~~}r@{~~}l|l@{~~}r|l@{~~}r|l@{~~}r|l@{~~}r|l@{~~}r|l@{~~}r@{~~}}
    a/h & Model & $w^{*}$ & \% & $\sigma_{13}(B)$ & \% & $\sigma_{23}(A)$ & \% & $\sigma_{23}(B)$ & \% & $\sigma_{13}(A)$ & \% & $\omega^{*}$ & \% \\ 

    \hline
    2 & ToSDT       & $4.7092$ & $-8.64$  & $1.9616$ & $+12.37$ & $2.2623$ & $-13.82$ & $1.0600$ & $+304.25$  & $0.94245$ & $+126.05$ & $4.2318$ & $+6.05$ \\ 
  & \tozzfour{} & $4.9169$ & $-4.61$  & $1.5936$ & $-8.71$  & $2.7553$ & $+4.96$  & $0.49673$ & $+89.43$  & $0.45655$ & $+9.51$   & $4.1849$ & $+4.87$ \\ 
  & \sizzfour{} & $5.1770$ & $+0.44$  & $1.5757$ & $-9.73$  & $2.5427$ & $-3.14$  & $0.51878$ & $+97.84$  & $0.49927$ & $+19.75$  & $4.0188$ & $+0.71$ \\ 
  & \tdfourwf{} & $5.4512$ & $+5.76$  & $1.8548$ & $+6.26$  & $2.8421$ & $+8.27$  & $0.37480$ & $+42.93$  & $0.48120$ & $+15.42$  & $4.0041$ & $+0.34$ \\ 
  & EToSDT      & $4.4825$ & $-13.04$ & $1.9471$ & $+11.54$ & $2.2050$ & $-16.00$ & $0.94772$ & $+261.41$ & $0.83878$ & $+101.19$ & $4.2472$ & $+6.43$ \\ 
  & \tozzfive{} & $4.6991$ & $-8.83$  & $1.5661$ & $-10.28$ & $2.6994$ & $+2.83$  & $0.36401$ & $+38.82$  & $0.39793$ & $-4.55$   & $4.2069$ & $+5.42$ \\ 
  & \sizzfive{} & $4.9483$ & $-4.00$  & $1.5538$ & $-10.99$ & $2.4876$ & $-5.24$  & $0.39432$ & $+50.37$  & $0.43174$ & $+3.56$   & $4.0409$ & $+1.26$ \\ 
  & \tdfivewf{} & $5.1481$ & $-0.12$  & $1.7458$ & $+0.01$  & $2.6239$ & $-0.05$  & $0.22315$ & $-14.90$  & $0.38768$ & $-7.01$   & $3.9907$ & $+0.00$ \\ 
  & Exact & $5.1544$ &  & $1.7456$ &  & $2.6252$ &  & $0.26223$ &  & $0.41692$ &  & $3.9905$ &  \\ 

	\end{tabular}
	\normalsize
	\caption{Comparison between the different models for the square $[-45/0/45/90]_s$ composite plate with $a/h=2$.}
	\label{tab:m45p0p45p90s}
\end{table*}
\section{Conclusion}
In this paper, a multilayered equivalent-single-layer plate theory with normal deformation has been presented. This theory, named 3D-5WF, is based on the use of five \emph{warping functions} \WFfirst{}, four of them describing the transverse shear behaviour and the fifth describing the normal deformation. The five \WFs{} are issued from 3D exact solutions of the bending (or of the dynamic response) of the simply supported laminate under bi-sine load. Hence the five \WFs{} depend on the lamination sequence, on the length-to-thickness ratio, and on the frequency. This leads to an adaptable theory which is able to give precise results for various laminates including single layer and sandwich plates, regardless of the length-to-thickness ratio which has been lowered up to 2 in this study.
\par
The present theory is compared to other theories and to exact solutions. Theories for comparison are of two kinds: i) models without normal deformation: the third order shear deformation theory (ToSDT, Reddy), a third order zig-zag theory (ToZZ4, Cho--Parmerter), a sine zig-zag theory (SiZZ4, enhancement of the Beakou--Touratier theory), and a theory based on \WFs{} issued from 3D solutions (3D-4WF, Loredo--Castel), and ii) enhancements of previous theories allowing normal deformation, leading to the EToSDT, ToZZ5, SiZZ5 and the 3D-5WF theories. All these theories have been formulated with the help of \WFs{}, so the solution procedure is unique. The problem which is solved for comparison is the simply supported plate with bi-sine load, for which exact solutions are known. The comparisons are made on deflections, stresses and fundamental frequencies for length-to-thickness ratio varying from~$2$ to~$10$. 
\par
Five lamination sequences are considered. Three single-ply plates are studied, the $[iso]$ one, made up with an isotropic material, and the $[0]$ and $[5]$ ones, made up with an orthotropic composite material. Theories are also compared for the study of a $[0/c/0]$ sandwich plate and a $[-45/0/45/90]_s$ symmetric angle-ply plate.
\par
The study of the $[iso]$ laminate shows that, for very low length-to-thickness ratios, the theories without normal deformation are not pertinent, compared to their equivalent with normal deformation.
\par
The EToSDT theory is not material-dependent as it is based on a cubic \WF{} for the shear behaviour and on a quadratic one for the normal deformation. The considered zig-zag theories (ToZZ5 and SiZZ5) are not material-dependent when they are used for a single layer plate. The study of the $[0]$ laminate shows that the ToZZ5 and SiZZ5 theories cannot adapt themselves to the different shear/longitudinal modulus ratio in the $x$ and $y$ directions, in other words $\varphi_{11}(z)=\varphi_{22}(z)$.
\par
The study of the $[5]$ laminate shows that, as the zig-zag mechanism is inoperative on a one-layer laminate, both ToZZ5 and SiZZ5 theories give null $\varphi_{12}(z)$ or $\varphi_{21}(z)$ \WFs{}, while they are scheduled to propose non null $\varphi_{12}(z)$ or $\varphi_{21}(z)$ functions on multilayered angle-ply laminates. On the contrary, the present model have non null cross \WFs{} which better describes the reality.
\par
The studies of the $[0/c/0]$ sandwich panel and of the $[-45/0/45/90]_s$ laminate, show that, as expected, ZZ theories differentiate themselves from the ToSDT when multilayered structures are considered. This is particularly evident when the transverse stresses of the $[-45/0/45/90]_s$ plate are computed. The two considered ZZ models (\tozzfive{} and \sizzfive{}) have four shear \WFs{} which permit them to consider a kinematic field that respect the transverse stress continuity at each interface for angle-ply structures.
\par
All considered theories could have given better results if higher length-to-thickness ratios had been considered. Low length-to-thick\-ness ratios (2, 4) can be considered as unrealistic, but for dynamic analysis, the effective length to thickness ratio depends on the wavelength, then can reach such low values. Comparisons show that the present theory gives better results than other tested theories, for all considered lamination sequences, and for all length-to-thickness ratios. However, this result has to be seen in the special context of this study. Plate problems solved in this study are simply supported problems with bi-sine loading. The \WFs{} used for the present theory are issued from an exact 3D solution of this particular bending problem. Their shapes depend on the lamination sequence and on the length-to thickness ratio. Although it can be expected that these \WFs{} will perform better than explicit cubic or sine ZZ \WFs{} in practical cases involving various boundary conditions, geometry, and loads, results may be not as good as than those of the academic case. One can say that for studies of any sort, the local bending and the local effective length-to-thickness ratio imply different shear behaviours, hence that the \WFs{} should be defined locally. Further studies need to be done in order to validate or adapt the process.
%

%
%
\bibliographystyle{model3-num-names}
\bibliography{ExtMultPlateTheory}
\appendix
%
%
%
\label{lastpage}
\end{document}